\documentclass[sigconf]{acmart}

\usepackage[]{inputenc}
\usepackage{graphicx}
\usepackage{epstopdf}
\usepackage{listings}
\usepackage{color}
\definecolor{keywordcolor}{rgb}{0.8,0.1,0.5}
\definecolor{webgreen}{rgb}{0,.5,0}
\usepackage[nice]{nicefrac}
\usepackage{epsfig}
\usepackage{graphicx}
\graphicspath{{./figs/}}
\usepackage{booktabs}
\usepackage{amsmath}
\usepackage{datetime}
\usepackage{bbm}
\usepackage{wasysym}
\usepackage{balance}
\usepackage{color,marvosym,xspace}
\usepackage{hyperref}
\usepackage{url}
\usepackage{multirow}
\usepackage{listings}
\usepackage{diagbox}
\usepackage{units}
\usepackage{subfigure}
\usepackage{enumitem}
\usepackage{ifsym}
\usepackage{lettrine}
\usepackage{relsize}
\usepackage{tikz}
\usepackage{pifont}
\usepackage{natbib}
\usepackage{soul}
\usepackage{subfigure}
\usepackage{algpseudocode}
\usepackage{makecell}
\usepackage{adjustbox}
\usepackage{stfloats}
\usepackage[normalem]{ulem}
\usepackage[ruled]{algorithm2e}
\usepackage{amsthm}

\makeatletter \g@addto@macro{\UrlBreaks}{\do\/\do\-} \makeatother

\let\sigproof\proof\let\proof\relax
\let\sigendproof\endproof\let\endproof\relax

\SetAlFnt{\small}
\SetAlCapFnt{\small}
\SetAlCapNameFnt{\small}
\SetAlCapHSkip{0pt}
\IncMargin{-\parindent}

\let\proof\sigproof
\let\endproof\sigendproof

\usepackage{booktabs}
\newtheoremstyle{sig}
  {}
  {}
  {\itshape}
  {}
  {\scshape}
  {.}
  {.5em}
  {#1 #2\thmnote{\quad(#3)}}

\makeatletter

\newcommand{\Rmnum}[1]{\expandafter\@slowromancap\romannumeral #1@}

\theoremstyle{sig}

\newcommand{\old}[1]{}

\setcopyright{acmcopyright}
\copyrightyear{2021}
\acmYear{2021}
\acmDOI{10.1145/1122445.1122456}

\acmConference[CCS 2021]{The 28th ACM Conference on Computer and Communications Security}{14 - 21 November, 2021}{Seoul, South Korea}
\acmBooktitle{2021 ACM SIGSAC Conference on Computer and Communications Security (CCS ’21), November 14–19, 2021, Virtual Event,
South Korea}
\acmPrice{15.00}
\acmISBN{978-1-4503-XXXX-X/18/06}

\makeatletter
\renewcommand\@formatdoi[1]{\ignorespaces}
\makeatother

\AtBeginDocument{%
  \providecommand\BibTeX{{%
    \normalfont B\kern-0.5em{\scshape i\kern-0.25em b}\kern-0.8em\TeX}}}

\begin{document}

\title{TableGAN-MCA: Evaluating Membership Collisions of GAN-Synthesized Tabular Data Releasing} 
\author{Aoting Hu}
\affiliation{%
  \institution{Southeast University}
  \country{China}
}

\author{Renjie Xie}
\affiliation{%
  \institution{Southeast University}
  \country{China}
}  

\author{Zhigang Lu}
\affiliation{
	\institution{Macquarie University}
	\country{Australia}
}

\author{Aiqun Hu}
\affiliation{%
  \institution{Southeast University}
  \country{China}
  }

\author{Minhui Xue}
\affiliation{
\institution{The University of Adelaide}
\country{Australia}
}

\pagestyle{plain}

\begin{abstract}
Generative Adversarial Networks (GAN)-synthesized table publishing lets people privately learn insights without access to the private table. However, existing studies on Membership Inference (MI) Attacks show promising results on disclosing membership of training datasets of GAN-synthesized tables. Different from those works focusing on discovering membership of a given data point, in this paper, we propose a novel Membership Collision Attack against GANs (\emph{TableGAN-MCA}), which allows an adversary given only synthetic entries randomly sampled from a black-box generator to recover partial GAN training data. Namely, a GAN-synthesized table immune to state-of-the-art MI attacks is vulnerable to the \emph{TableGAN-MCA}. The success of \emph{TableGAN-MCA} is boosted by  an observation that GAN-synthesized tables potentially collide with the training data of the generator.

Our experimental evaluations on \emph{TableGAN-MCA} have five main findings. First, \emph{TableGAN-MCA} has a satisfying training data recovery rate on three commonly used real-world datasets against four generative models. Second, factors, including the size of GAN training data, GAN training epochs and the number of synthetic samples available to the adversary, are positively correlated to the success of \emph{TableGAN-MCA}. Third, highly frequent data points have high risks of being recovered by \emph{TableGAN-MCA}. Fourth, some unique data are exposed to unexpected high recovery risks in \emph{TableGAN-MCA}, which may attribute to GAN's generalization. Fifth, as expected, differential privacy, without the consideration of the correlations between features, does not show commendable mitigation effect against the \emph{TableGAN-MCA}. Finally, we propose two mitigation methods and show promising privacy and utility trade-offs when protecting against \emph{TableGAN-MCA}.

\end{abstract}



\maketitle

\section{Introduction}
\begin{figure}
  \centering
  \includegraphics[width=\linewidth]{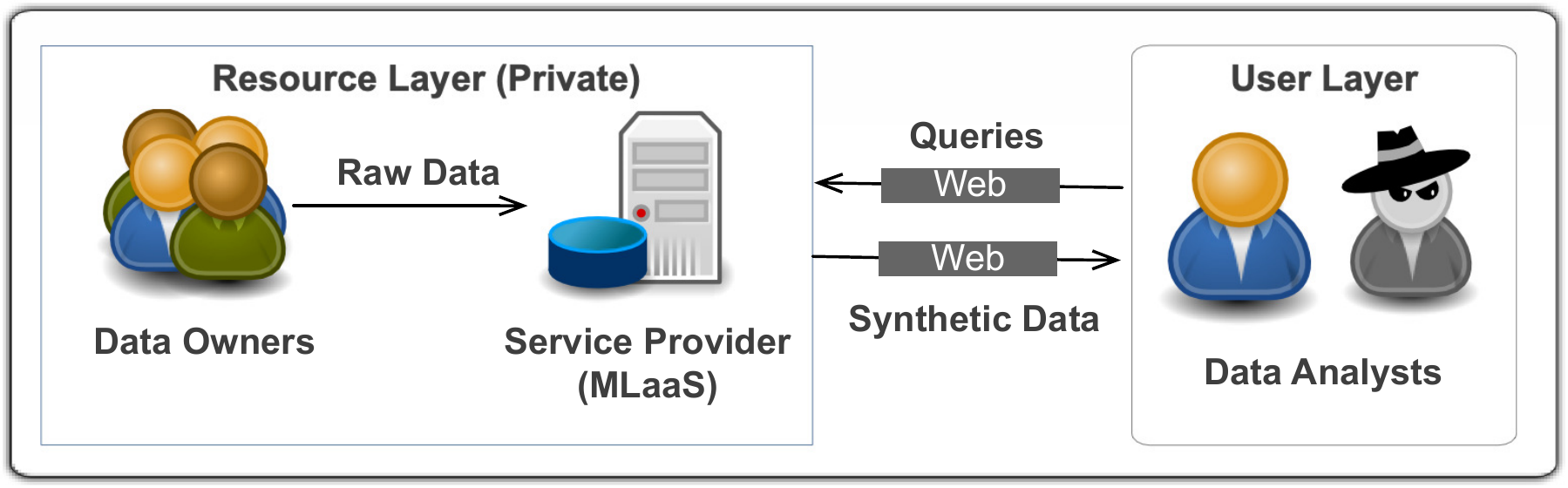}
  \vspace{-4mm}
  \caption{The framework of private data publishing. Both the data owner and service provider who guard resources are trusted. The data analysts are legal customers as well as potential adversaries.}
  \label{fig:frame}
\end{figure}

Big data have emerged as valuable resources that allow companies, researchers and governments to enhance decision making, insight discovery and process optimization. However,
sharing sensitive datasets without violating individual's privacy is a long-standing challenge. 
For example, in $2017$, DeepMind was accused of an illegal acquisition of personal medical records of $1.6$ million patients for developing a kidney injuries diagnosing application~\cite{Google}. To analyze those sensitive data in a privacy-preserving manner, ideally, we need a trusted third party that collects and processes raw data, and then releases a sanitized version of data trading off privacy and utility through web queries (see the paradigm shown in Fig.~\ref{fig:frame}). 

However, state-of-the-art solutions for releasing the sanitized data  achieving trade-offs between utility and privacy are vulnerable to privacy inference attacks. For example, de-identification (removing unique identifiers for all data entries) is susceptible to linkage attacks~\cite{Narayanan2008}.  Anonymization~\cite{sweeney2002k,machanavajjhala2006diversity,li_t-closeness_nodate} suffers from background information attacks. Other synthetic dataset publishing mechanisms, such as NetMechanism~\cite{blum2013learning}, Iterative Construction~\cite{hutchison_iterative_2012,gupta_privately_nodate,hardt_multiplicative_2010,hardt_simple_nodate}, are tailored for relatively small datasets~\cite{dwork2014algorithmic}. 
More recently, Generative Networks, including Generative Adversarial Networks (GANs)~\cite{goodfellow2014generative} and Variational Autoencoders (VAEs)~\cite{kingma2013auto}, produce synthetic data that achieve  enhanced privacy and utility trade-offs. Such synthetic data conceal the detailed  (privacy) of the raw data while keeping statistics similarity~\cite{xu2019modeling,park2018data}. Nevertheless, recent works~\cite{park2018data,hayes_logan_2019,hilprecht_monte_2019,chen_gan-leaks_2020,stadler_synthetic_2020} show the risk of membership disclosure (i.e., inferring whether a given data point belongs to the training dataset) against synthesized data by attacking generator APIs. They propose various Membership Inference Attacks (MIAs) against published generative models to disclose the membership information of training data.

To further explore the privacy disclosure risks of the GAN-synthesized tabular data, different from existing MIAs against generative models~\cite{park2018data,hayes_logan_2019,hilprecht_monte_2019,chen_gan-leaks_2020,stadler_synthetic_2020}, we propose a novel attack model, named Membership Collision Attack against GAN-synthesized Tables (\emph{TableGAN-MCA}). Specifically, we reconstruct a proportion of actual training data from the published synthetic table with high confidence by inferring the membership collisions (substantiated in Section~\ref{sec:MCP}). Hence, \emph{TableGAN-MCA} brings a novel privacy problem: training data exposure when analyzing published synthetic tabular data. In addition, \emph{TableGAN-MCA} only queries a black-box generator (of the GAN) for synthetic data, which is similar to the most strict threat model introduced in the recent work - GAN-Leaks~\cite{chen_gan-leaks_2020}. We conceptualize the differences among recent works in Table~\ref{tab:compare}.

\noindent \textbf{Motivation.} Our work is motivated by two observations in GAN-synthesized table (low-dimensional data) releasing. 

\begin{figure}[t]
  \centering
  \includegraphics[width=0.7\linewidth]{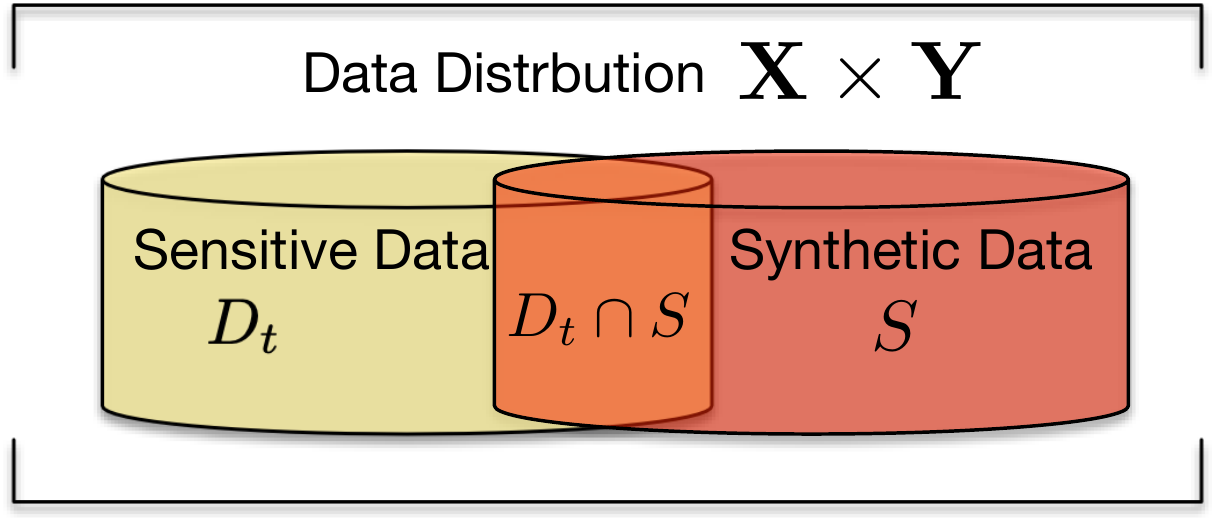}
  \vspace{-1mm}
  \caption{The training dataset $D_t$ intersects the synthetic dataset $S$ at $D_t \cap S$.}
  \label{fig:venn2}
\end{figure}

\begin{itemize}[leftmargin=*]
\item \textbf{Observation~1.} Generated synthetic tables overlap with GAN's training data (as the intersection illustrated in  Fig.~\ref{fig:venn2}). For instance, in the Adult dataset, a synthetic dataset collides with the GAN's training dataset by $16.9\%$ ($5350$ entries). Clearly, such an overlap brings severe privacy breaches if adversaries could locate the intersection. In the remainder of this paper, we call the overlap/intersection \textit{membership collision}.

\item \textbf{Observation~2.} In the GAN-synthesized tabular data, membership collisions and data frequency are positively correlated (substantiated in Fig.~\ref{fig:observation}).  
However, it is rare to trigger sample collisions in high-dimensional data, such as image synthesis, due to the curse of dimensionality. Thus, the distribution of tabular data with relatively small dimension 
brings additional privacy risks than that of image synthesis.  
\end{itemize}

To perform the proposed \emph{TableGAN-MCA}, we leverage shadow models~\cite{shokri2017membership} to learn the patterns behind the collision (\textit{Observation~1}) while taking the density of each synthetic data by counting its sample frequency in synthetic distribution (\textit{Observation~2}) as additional feature when training the attack model. \emph{TableGAN-MCA} shows promising results on commonly used real-world datasets, including Adult, Lawschool and Compas. For instance, \textbf{\emph{TableGAN-MCA} recovers $36.1\%$, $12.7\%$, $36.5\%$ of actual members released with the GAN-synthesized tabular data with approximately 80\% confidence for Adult, Lawschool and Compas, respectively.} Our results show that a well-trained GAN, robust to the MIAs proposed in~\cite{park2018data,hayes_logan_2019, chen_gan-leaks_2020}, is still vulnerable to \emph{TableGAN-MCA}.

In summary, our main contributions are as follows: 
\begin{itemize}[leftmargin=*]

\item 
We propose a novel membership collision attack against GAN-synthesized tabular data publishing, named \emph{TableGAN-MCA}, which can reinstate partial training data with high confidence. \emph{TableGAN-MCA} exploits the weaknesses of GAN synthesis observed on low-dimensional data, i.e., GAN-synthesized data collide with its training data, and members (in the colliding member set) occur more frequently than non-members.

\item We extensively evaluate our proposed attacks on three commonly used real-world datasets, including Adult, Lawschool and Compas  against four generative models, including TVAE~\cite{xu2019modeling}, CTGAN~\cite{xu2019modeling}, WGAN-GP~\cite{gulrajani2017improved} and WGAN-WC~\cite{arjovsky2017wasserstein}.
Furthermore, we explore the factors that may impact the attack effectiveness, such as the size of GAN training data, GAN training epochs, GAN training data frequencies and the number of synthetic samples available to the attacker. 

\item We discover that individuals in the training dataset have various risks of privacy leakage under \emph{TableGAN-MCA}. Additionally, we show that GANs do not memorize those exposed data. Instead, when generalizing the distribution of the training data, GANs may increase or decrease the frequency of some individuals, and hence change their privacy risks. 
 
\item We examine the effect of differential privacy (DP) to mitigate \emph{TableGAN-MCA}. Our empirical results show that differential private generative model training achieves sub-optimal trade-offs against \emph{TableGAN-MCA}. It is mainly due to the fact that \emph{TableGAN-MCA} relies more on the common pattern of a distribution (like attribute correlations) which is not the focus of DP. In addition to DP, we propose two mitigation methods, naive defense and improved defense, that mitigate the effect of \emph{TableGAN-MCA}. 
 
\end{itemize}

\begin{table}[!t]
	\centering
	\caption{Comparison with MIAs against GANs. ($\blacksquare$: black-box access;  --: insufficient information provided; $\surd$: require; $\times$: does not require)}
	   \vspace{-1mm}
	\label{tab:compare}%
	\scalebox{0.8}{
	\begin{tabular}{ccccccc}
		\toprule
		& \makecell[c]{Benchmark\\ Datasets}  &  \makecell[c]{$\blacksquare$ Gen-\\erator} & \makecell[c]{$\blacksquare$ Dis-\\criminator} & \makecell[c]{Extra\\Targets}  &  \makecell[c]{Expose\\Trainset}\\
		\midrule
		LOGAN~\cite{hayes_logan_2019} & Image & $\surd$ & $\surd$   & $\surd$     & False   \\
		table-GAN~\cite{park2018data} & Table  & $\surd$ & $\surd$   & $\surd$     & False   \\
		MC~\cite{hilprecht_monte_2019} & Image & $\surd$ & $\times$   & $\surd$   & --  \\
		GAN-leaks~\cite{chen_gan-leaks_2020} & Image/Table & $\surd$ & $\times$  & $\surd$     &  False \\
		\textbf{TableGAN-MCA} & Table & $\surd$ & $\times$ & $\times$  & True \\		
		\bottomrule
	\end{tabular}%
	}
\end{table}%

\section{Background of Generative Models}
\label{backgound}
Generative Adversarial Networks (GANs)~\cite{goodfellow2014generative} and its variants have made great achievements in generating high quality artificial data that mimic the real ones, by modeling the underlying data distribution. It is composed of two neural networks: a discriminator $D$ and a generator $G$. It tries to minimize the distance between the real data distribution $\mathbf{P}_r$ and the generated (artificial) data distribution~$\mathbf{P}_{g}$ by iteratively updating parameters of the networks.

The Wasserstein GAN (WGAN)~\cite{arjovsky2017wasserstein} applies Earth Mover (EM) distance under a K-Lipschitz constraint and achieves good performance in generating high fidelity samples. The loss function of the discriminator and the generator are as follows: 
\vspace{-1mm}
\begin{equation}
J^{(D)}\left(\boldsymbol{\theta}^{(D)}, \boldsymbol{\theta}^{(G)}\right)=-\frac{1}{2} \mathbb{E}_{\boldsymbol{x} \sim p_{\mathrm{data}}} D(\boldsymbol{x})+\frac{1}{2} \mathbb{E}_{\boldsymbol{z}} D(G(z)),
\end{equation}
\vspace{-2mm}
\begin{equation}
J^{(G)}=-\frac{1}{2} \mathbb{E}_{z} D(G(z)).
\end{equation}

In this work, we use its weight clipping version (WGAN-WC)~\cite{arjovsky2017wasserstein},  Gradient Penalty version (WGAN-GP)~\cite{gulrajani2017improved} and CTGAN (state-of-the-art)~\cite{xu2019modeling}. We also include TVAE from ~\cite{xu2019modeling} for its comparable performance as CTGAN. Following \cite{xu2019modeling}, all three GANs uses recurrent networks in the generator. For categorical features, we use the gumble-softmax activation in the output of the generator. For numerical features, we use the sigmoid or the tanh activation in the output of the generator based on value range. The architecture and parameters of GANs are broken down in Appendix~\ref{app_a}.

\section{Problem Formulation} 
In this section, we formulate our membership collisions problem, followed by the description of the threat model according to adversary's goals, capabilities and background knowledge. We introduce all the notations used throughout the paper in Table~\ref{tab:Notations}. 

\begin{table}
	\centering
	\caption{Summary of notations.}
	\label{tab:Notations}
	\scalebox{0.9}{
	\begin{tabular}{clcl}
		\toprule
		Symbol & Description & Symbol & Description  \\
		\midrule
		$D_t$ & Private training dataset &       
		$D_s$ & Test dataset \\
		$S$ & Released synthetic dataset &
		$\widetilde{S}$ & Shadow dataset \\
		$\mathbf{P}_r$ & Training data distribution & $G$ & Generator oracle \\
		$\mathbf{P}_z$ & Prior Gaussian distribution & $\mathbbm{1}$ & Indicator function \\
		$\mathbf{P}_g$ & Generated data distribution &
		$\mathbf{x}$ & A data point \\
		$I$ & colliding member set &
		$\mathcal{A}$ & Adversary \\
		$N_s$  &   Number of synthetic copies  & $f(\cdot)$ & Attack classifier \\
		\bottomrule
	\end{tabular}%
	}%
\end{table}%

\subsection{Membership Collision Problem} 
\label{sec:MCP}

We let $D_t =\{\mathbf{x}\}$ be a training set sampled from an implicit data distribution $\mathbf{P}_r$. Each private entry takes the form as $\mathbf{x}=(x, y) \in \mathbf{X} \times \mathbf{Y}$, where $x$ represents the features and $y$ represents the class label. A data release mechanism GAN trains on the training set $D_t$ and outputs a well learned generator $G$. Generator $G$ is a deterministic function that maps a prior distribution, i.e., Gaussian distribution $\mathbf{P}_z$, to the generated distribution $\mathbf{P}_g$ that mimic real distribution $\mathbf{P}_r$. Then, a synthetic dataset $S \sim \mathbf{P}_g$ is published and serves as a sanitized version of $D_t$. We formalize the \textbf{membership collisions} as : a published synthetic datasets $S \sim \mathbf{P}_g$ collide with its training set $D_{t} \sim \mathbf{P}_r$ and result in a colliding member set $I = S \cap D_t$. Notice that a data point $\mathbf{x}\in I$ result in $\mathbf{x} \in D_t$. Similarly, a synthetic data point $\mathbf{x} \not\in I$ result in $\mathbf{x} \not\in D_t$.

We aim to study how much an adversary~$\mathcal{A}$ increases its ability to assert whether a synthetic data point $\mathbf{x}\sim S$ belongs to the colliding member set $I$ by estimating the generated distribution $\mathbf{P}_g$ via the published synthetic dataset $S$. Formally, 
\begin{definition}[Membership Collision Attack]
\label{MCIA}
	Given a synthetic dataset $S$ produced by a generative model $G(\mathbf{P}_z,D_t)$ that contains a colliding member set $I = S \cap D_t$ and an attack algorithm $\mathcal{A}(\mathbf{x})$ that outputs $1$ if it outputs the synthetic data $\mathbf{x} \in I$, we say the generative model $G$ is subject to membership collision inference attack if there exists an entry $\mathbf{x} \in S$ such that 
	\begin{equation}
		\Pr[\mathcal{A}(\mathbf{x},\mathbf{P}_{g}) = 1 ]-\Pr[\mathcal{A}(\mathbf{x}) = 1] > \alpha,
	\end{equation}
	where $\alpha$ is a non-negligible value.
\end{definition}

In this work, we consider that the prior advantage of the attacker is random guess, that is, $\Pr[\mathcal{A}(\mathbf{x}) = 1] = \Pr[\mathbf{x}\in I]$. Thus, we evaluate the posterior advantage of the attacker thereafter. 

Note that Def.~\ref{MCIA} differs from the membership inference definition~\cite{shokri2017membership} by changing the goal of arbitrary membership inference with membership collision inference of synthetic data.  The proposed \emph{TableGAN-MIA} is an instance of MCA in GAN-synthesized table releasing.

\subsection{Threat Model}  
\label{sec:threat_model}
In the context of GAN-synthesized data sharing, adversaries are external parties that wish to learn the statistics of the sensitive dataset by querying data owners or curators. 
In existing MIAs against GANs~\cite{park2018data,hayes_logan_2019,hilprecht_monte_2019,chen_gan-leaks_2020}, the adversary's knowledge is: (1) having only limited synthetic data, (2) accessing a black-box generator API (unlimited synthetic data), (3) accessing a black-box generator plus a discriminator oracle, (4) accessing a white-box GAN. Our study focus on the most strict attack model: (1) and (2) (which is similar to the threat model in MC~\cite{hilprecht_monte_2019} and ``Full Black-box Generator" assumption in GAN-Leaks~\cite{chen_gan-leaks_2020}). The attacker does not know the priori of the model's structure, including meta-parameters, training data and any target data to infer membership. In \emph{TableGAN-MCA}, the adversary's goal is to recover the value of some members of the training set from the published synthetic datasets that may unintentionally contain colliding members. In this paper, we evaluate \emph{TableGAN-MCA} under two threat models:

\noindent \textbf{Attack model (1)}: accessible to limited synthetic data. We assume the adversary has one copy of synthetic dataset $S$ following $\mathbf{P}_{g}$, of size $|S|=|D_{t}|=n$.

\noindent \textbf{Attack model (2)}: accessible to unlimited synthetic data. We assume the adversary has $N_s$ ($N_{s}$ is a positive integer) synthetic copies $\{S_1,S_2, \ldots, S_{N_s}\}$, each of which has size $|S_{i}|=|D_{t}|=n$.

\begin{figure}
	\centering 
	\includegraphics[width=\columnwidth]{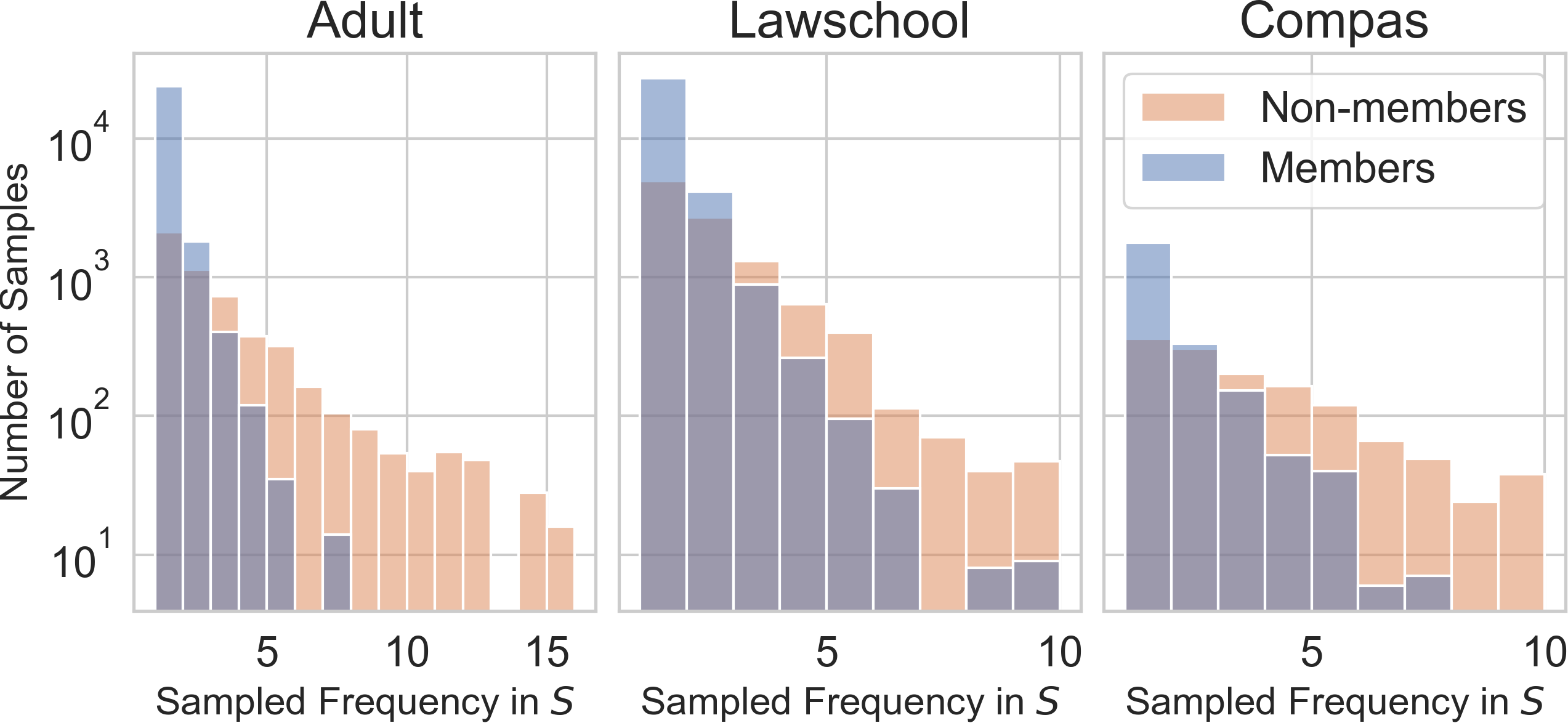}%
	\label{ECDF}
\vspace{-2mm}
\caption{Comparisons of sample frequency between members and non-members. The x-axis represents all possibilities of $\{\# \mathbf{x}_i\}$ in published synthetic datasets and y-axis represents log of the number of eligible data points.}
\Description{The x-axis represents all possibilities of $\{\# x_i\}$ and y-axis represents log of the number of eligible data points.}
\vspace{4mm}
\label{fig:observation}	

\end{figure}

\section{Membership Recovery Framework against GAN-synthesized Tables}

In this section, we propose a membership indicator for inferring membership collisions from the statistics of the published table. Based on the membership indicator, we propose the \emph{TableGAN-MCA} to recover the value of the training set of GAN-synthesized tables in the black-box setting.

\vspace{-2mm}
\subsection{Membership Indicator}
The membership indicator is triggered by two observations. First, the released GAN-synthesized tables often overlap the training dataset of the GAN model. Second, such synthetic data points appearing frequently in the published GAN-synthesized data are more likely to be the colliding member of the training dataset. That is, $\Pr[\mathbf{x}\in D_t|\mathbf{P}_g] \propto \Pr[\mathbf{x}|\mathbf{P}_g]$. Fig.~\ref{fig:observation} depicts the observations from three datasets used in this paper, where we count the numbers of members and non-members, given numbers of appearance of the data points in the released synthetic tables. In Fig.~\ref{fig:observation} (left), approximately $96\%$ of synthetic data with a sampled frequency of more than three are colliding members. Conversely, almost $91\%$ unique synthetic data are non-colliding members in the Adult dataset. Thus, sample frequency is highly correlated with membership collisions and can be treated as an indicator to indicate membership. Formally, we estimated the membership indicator by the following equation.

\begin{equation}
\label{eq:prob}
    \operatorname{Pr}\left[\mathbf{x}_{i} \mid P_{g}\right] \approx  \underset{\mathbf{x}_{j} \in S}{\mathbb{E}} \mathbbm{1} \left(\mathbf{x}_{i}=\mathbf{x}_{j}\right) = \frac{1}{n} \sum_{j=1}^{n} \mathbbm{1} \left(\mathbf{x}_{i} = \mathbf{x}_{j}\right),
\end{equation}
where an indicator function $\mathbbm{1}(\cdot)$ outputs $1$ if its argument is true, $S$ is the synthetic datasets available to the adversary following $\mathbf{P}_g$, of size $|S|=n$.

To date, the adversary can launch a data reconstruction attack by setting a threshold for the value of a membership collisions indicator of Eq.~\eqref{eq:prob}, similar to~\cite{chen_gan-leaks_2020}. The adversary then claims that the synthetic data, having collisions indicators greater than a given threshold, are the recovered data. However, choosing an optimal threshold is a non-trivial task for an adversary without background knowledge about training data except the published synthetic data. 
To deal with it, we additionally leverage shadow model techniques~\cite{shokri2017membership} to enhance the knowledge of adversaries to construct a robust \emph{TableGAN-MCA} framework.

\subsection{TableGAN-MCA}
\begin{figure}[t]
	\centering  
	\includegraphics[width=\columnwidth]{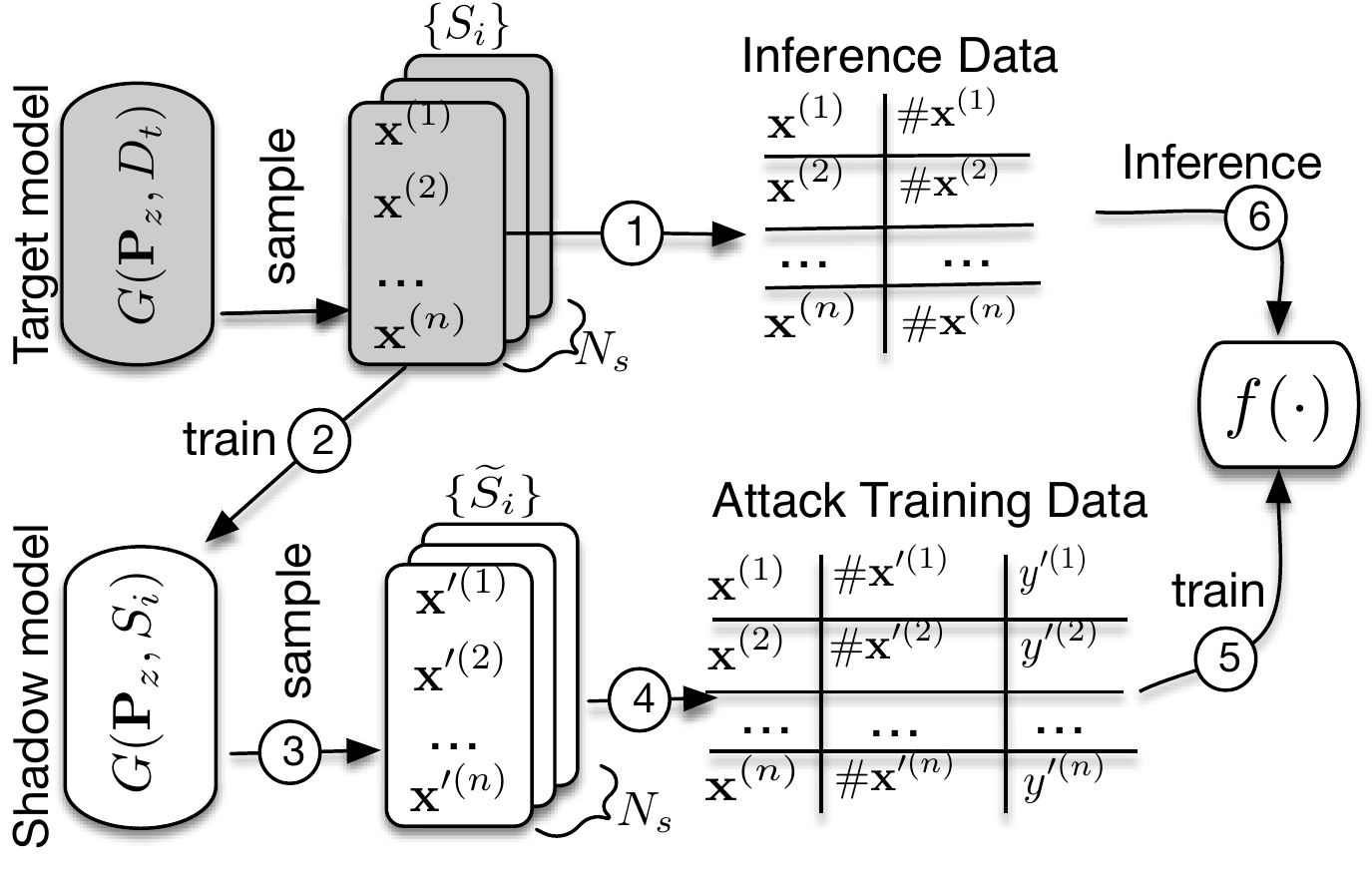}%
	\vspace{-1mm}
	\caption{The overview of the procedures of \emph{TableGAN-MCA} against the black-box generator in data synthesis.}
\label{fig:adv}	

\end{figure}

In a nutshell, \emph{TableGAN-MCA} combines the  membership collisions indicator and the shadow models ~\cite{shokri2017membership} to train an attack model to learn the relation between membership collisions (labels) and indicator values (features) in released GAN-synthesized tables. Fig.~\ref{fig:adv} depicts the framework of \emph{TableGAN-MCA} and Alg.~\ref{alg:TableGAN-MCA} shows the detailed implementation. Each step in Alg.~\ref{alg:TableGAN-MCA} corresponds to the step index in Fig.~\ref{fig:adv}. In summary, steps 2, 3, 4 and 5 train an attack classifier by giving synthetic data. Steps 1 and 6 infer membership collisions to recover training data. 

In Steps $1$ and $4$, $\{\#\mathbf{x}\}$ represents estimated sample frequency following from Eq.~\ref{eq:prob}. They are concatenated (``$\bowtie$'') to $S_i$ (Step 1) and $\widetilde{S}_{i}$ (Step 4) as an extra feature.

In Step 4, a label function is required to claim membership collisions in shadow datasets. For a shadow dataset $\widetilde{S}_i$ such that $S_i \cap \widetilde{S}_i = \widetilde{I}_i$, a membership collisions label for each data $\mathbf{x}^{\prime}_i$ will be 
$y^{\prime}_i=\mathbbm{1} \left(\mathbf{x}^{\prime}_i \in \widetilde{I}_i \right)$.

In Step $6$, attack model $f(\cdot)$ outputs the predicted probability about whether a synthetic data is colliding member. Adversaries then expose a data set $R_{\mathcal{A}}$ that with high prediction scores.

For attack model $(2)$ (unlimited synthetic data) such that $N_s>1$, the adversary repeat the Step1 to Step 4 $N_{s}$ times and gets $N_s$ labeled shadow datasets $\{\widetilde{S}_{1},\widetilde{S}_{2},...,\widetilde{S}_{N_s}\}$ such that each of them with size $N_{s} \times |D_t|$. Then the adversary concat (``$\Vert$'') all shadow datasets together to train the attack model.

\SetKwInput{KwInput}{Input}                
\SetKwInput{KwOutput}{Output} 

\begin{algorithm}[t]
\caption{TableGAN-MCA.}
\label{alg:TableGAN-MCA}
\DontPrintSemicolon
\KwInput{
$\{S_1, S_2,...,S_{N_s}\}$: Released synthetic datasets; $|D_{t}|$: Size of the training dataset;
}
\KwOutput{$R$: Recovered data from $D_{t}$}
\While {$i: 1 \to N_{s}$}
{
    \textbf{Step 1}: \\Frequency $\{\# \mathbf{x}_{i}\}$ $\gets$ Estimate frequency for each $\mathbf{x}_{i} \in S_{i}$ by Eq.~\eqref{eq:prob};\;
    $S_{i}$ $\gets$ $S_{i} \bowtie \{\# \mathbf{x}_{i}\}$;\;
    \textbf{Step 2}: Shadow GAN generator $\widetilde{G}_{i}$ $\gets$ Train on $S_{i}$;\;
    \textbf{Step 3}: Shadow set $\widetilde{S}_{i}$ $\gets$ Sample from $\widetilde{G}_{i}$, $|\widetilde{S}_{i}| = N_{s} \times |D_{t}|$.\;
    \textbf{Step 4}: \\Frequency $\{\# \mathbf{x}^{\prime}_{i}\}$ $\gets$ Count the frequency for each $\mathbf{x}^{\prime}_{i} \in \widetilde{S}_{i}$ by Eq.~\eqref{eq:prob};\;
    $\widetilde{S}_{i}$ $\gets$ $\widetilde{S}_{i} \bowtie \{\# \mathbf{x}^{\prime}_{i}\}$;\;
    Ground truth label $y^{\prime}_i$ $\gets$ $
\mathbbm{1} \left(\mathbf{x}^{\prime}_i \in \widetilde{I}_i \right)$;\;
    }
\textbf{Step 5}: 
TableGAN-MCA attack model $f(\cdot)$ $\gets$ Train on $\big\Vert_{i=1}^{N_s} \widetilde{S}_{i} $ with member/non-member labels $\big\Vert_{i=1}^{N_s} \{y^{\prime}_i\}$, where $\big\Vert_{i=1}^{N_s} \widetilde{S}_{i} = \widetilde{S}_{1} \Vert \dots \Vert \widetilde{S}_{N_s}$; $\big\Vert_{i=1}^{N_s} \{y^{\prime}_i\} = \{y^{\prime}_1\} \Vert \dots \Vert \{y^{\prime}_{N_s}\}$ ;\;
\textbf{Step 6}: $R_{\mathcal{A}}$ $\gets$ $f(\{S_i\})$;\;
\Return{$R_{\mathcal{A}}$}\;
\end{algorithm}

Note that in the worst-case (to the adversary), where the intersection between the training set and the synthetic dataset could be empty, the adversary of \emph{TableGAN-MCA} cannot recover anything from the private training data. To avoid such a case, we would discretize the synthetic dataset to generalize the range of each feature such that there is a non-empty intersection. In this way, we could (at least) recover coarse-grained information regarding the members within the training data. We show the details of the discretization operation in Section~\ref{sec:data}.
\section{Evaluation}
In this section, we first introduce the methods of tabular data synthesis, then introduce the evaluation metrics. Next we show the attack performance of \emph{TableGAN-MCA} as well as the comparisons with recent works.

\subsection{Dataset Synthesis}
\label{sec:data}
We perform experimental evaluations on three commonly used~\cite{ruoss_learning_2020,xu2019modeling,park2018data,barenstein_propublicas_2019,chen2018differentially} real-world tables, Adult~\cite{Adult}, Lawschool~\cite{Lawschool} and Compas~\cite{Compas}.

\noindent \textbf{Adult}: The US Adult Census dataset is a repository of $48842$ entries extracted from 1994 US Census dataset, where $45222$ entries have complete information. After pre-processing, it remains $1$ numerical feature, $12$ categorical features and $1$ binary label.

\noindent \textbf{Lawschool}: This dataset comes from the Law School Admission Council's National Longitudinal Bar Passage Study. It contains application records for $25$ different law schools with $86022$ individuals. It has $2$ numerical features, $5$ categorical features and $1$ binary label.   

\noindent \textbf{Compas}: COMPAS recidivism risk score and criminal history data is collected by ProPublica in 2016. After pre-processing, it remains $5278$ entries with $4$ numerical features, $6$ categorical features and $1$ binary labels.

Note that unlike MIAs attacking classifiers that produce predicted labels with probability, generative models only output synthetic samples. The labels in generated datasets serve as an ordinary feature like other features when training attack models. Therefore, for simplicity, we use the three binary-labeled datasets in our experiments.

\noindent \textbf{Tabular data synthesis.} For training generative models, we apply Tabular Variational Autoencoder (TVAE)~\cite{xu2019modeling}, CTGAN~\cite{xu2019modeling},  WGAN-GP~\cite{gulrajani2017improved} and WGAN-WC~\cite{arjovsky2017wasserstein} for their superior modeling quality in tabular synthesis. To facilitate data synthesis, we have the following additional data pre-processing. (1) We discretize  imbalanced and sparse numerical values in given columns to categorical values. (2) We normalize numerical columns into $(0,1)$ or $(-1,1)$. (3) We one-hot encode all categorical features (4) We split the dataset into the training set ($D_t$, $70\%$ records) and test set ($D_s$, $30\%$ records) (see row 1 and row 2 in Table~\ref{tab:dataset}). The training set is used for dataset synthesis and the test set is used for examining the utility of the synthetic data. 

\noindent \textbf{Discretization in pre-processing.} Features in tabular dataset are either categorical or numerical variables. Unlike pictures, some numerical columns are non-Gaussian distribution, that is, it either has long tails, sparse distribution or multiple modes. Generative models cannot model them well without appropriate pre-processing. 
To address this issue, we discretize the imbalanced and sparse numerical values to categorical values. In the experiments, such simple discretization in pre-processing exhibits decent performance in generating complex features while keeping original statistics. Note that discretization definitely makes some records of the original dataset share the same values (similar to $k$-anonymity~\cite{sweeney2002k}). We show the uniqueness of the records after pre-processing in Table~\ref{tab:dataset} (row three and four), where a large proportion of sensitive data points can still be uniquely identified before feeding into generative models.

\begin{table}[!t]
	\centering
	\caption{Dataset Statistics for GAN synthesis. $\operatorname{Pr}[\#\mathbf{x}=1]$: unique training data proportion; $\operatorname{Pr}[\#\mathbf{x} \leq 3]$: Proportion of training data with frequency less than $3$.}
	\vspace{-1mm}
	\label{tab:dataset}
	\begin{tabular}{ccccc}
		\toprule
		& Adult & Lawsch  & Compas \\
		\midrule
		$\#$ of Train $D_t$ ($70\%$) & 31655 & 60215  & 3694 \\        
		$\#$ of Test $D_s$ ($30\%$) & 13567 & 25807  & 1584 \\
		\hline
		$\operatorname{Pr}[\#\mathbf{x}=1]$ & $79.39\%$ &$71.28\%$ &  $63.72\%$  \\
		$\operatorname{Pr}[\#\mathbf{x} \leq 3]$ & $86.07\%$& $81.85\%$ & $74.28\%$\\
		\bottomrule
	\end{tabular}%
\end{table}%

\subsection{Metrics}

\subsubsection{Data Utility Metrics}
\label{utility_metrics}
For data utility evaluation, we consider two measurements: machine learning efficacy (models trained on a synthetic dataset and the original dataset provide similar predictions) and distribution fitness (a synthetic dataset is statistically similar to its original dataset in all attributes).  

For distribution fitness, we present $1$-way marginals that are approximated by the Empirical Cumulative Distribution Function (ECDF) for each attribute.
Having ECDFs of real and synthetic data, we compute attribute-wise Wasserstein distance, i.e., $ l_{1}(x_i,x^{\prime}_i) = \int_{-\infty}^{+\infty}|U_i-V_i| $, where $U_i$ and $V_i$ are respective CDFs of real attribute $x_i$ and synthetic attribute $x^{\prime}_i$~\cite{ramdas_wasserstein_2015}. We compare the expected value of ECDFs by $\mathbb{E}_i (l_{1}) = \frac{1}{n}\sum_{i=1}^{n}\{l_{1}(x_i,x^{\prime}_i)\}$.

\subsubsection{Attack Performance Metrics}

To evaluate the privacy of the released synthetic table, we consider membership collisions privacy, i.e., the \emph{TableGAN-MCA} effect. We use precision and recall to evaluate the attack performance (following Shokri et al.~\cite{shokri2017membership}), since the synthetic dataset that is used to inference has a skewed label distribution. Specifically, precision measures the probability of an entry inferred as a member is indeed the member of the training dataset, denoted as $\Pr (y=1|\hat{y}=1)$. Intuitively, it implies the confidence of the attacker in guessing positive membership. Recall measures the probability of a member is correctly inferred as a member by the attacker, denoted as $\Pr (\hat{y}=1|y=1)$. It reflects the percentage of positives exposed in the attack. In evaluation, we report precision and recall by Precision-Recall (PR) curve since it is more informative than ROC-curve under the case of skewed label distribution~\cite{davis_relationship_2006}. A higher Area under the PR-curve (AUPRC) implies both higher precision and recall, and thus they are used to compare the attack efficacy.

In addition to the attack precision and recall, we also consider a recovery rate because it reflects what the proportion of training data $D_t$ are being exposed to \emph{TableGAN-MCA}. Let $R_{\mathcal{A}}$ be recovered training data sets of the attack algorithm $\mathcal{A}$. The recovery rate $\rho_{\mathcal{A}}$ of $\mathcal{A}$ is defined as below:
\begin{equation}
\label{ep:recovery}
	\rho_{\mathcal{A}} =  |R_{\mathcal{A}}| / |D_t|.
\end{equation}
Note that the recovery rate shares the same numerator as the recall of the attack model $f(\cdot)$ but the different denominator ($|D_t|$ vs $|I|$).

\begin{table}\footnotesize
	\centering
	\caption{Model prediction accuracy ($\%$) trained on real training $D_t$ (``Base'') and GAN-synthesized datasets $S$. $\mathbb{E}_i (l_{1})$ denotes the average of all attribute-wise Wasserstein distance. }
	\vspace{-1mm}
	\label{tab:ML}%
	\begin{tabular}{c|c|cccc|c}
		\toprule
	     & Methods & DT & MLPC & Ada & LR & $\mathbb{E}_i (l_{1})$\\
		\midrule
		\multirow{5}{*}{\rotatebox[origin=c]{90}{Adult}} & Base & 85.39 &84.11 & 86.28 & 84.74 & 0 \\
		& TVAE & 79.24 & 77.53 & 78.72 & 80.2 & 0.0207 \\
		& CTGAN & 81.53 & 81.76 & 82.3 & 82.41 & 0.0266 \\
		& WGANWC & 82.74 & 83.62 & 84.16 & 83.96 & 0.0075\\
		& WGANGP & 83.16 & 83.95 & \textbf{84.24}& 83.93 & \textbf{0.0039} \\
		\midrule
		\multirow{5}{*}{\rotatebox[origin=c]{90}{Lawschool}}& Base & 81.90 & 89.54 & 87.23 & 87.68 & 0\\
		& TVAE & 79.53 & 85.38 & 85.17 & 85.26 & $0.0120$\\
		& CTGAN & 76.35 & 80.91 & 81.02 & 81.37 & 0.0283 \\
		& WGANWC & 77.54 & 80.76 & 80.56 &80.93 & 0.0073\\
		& WGANGP & 80.14 & 86.10 & 86.02 & \textbf{86.79} & \textbf{0.0047} \\
		
		\midrule
		\multirow{5}{*}{\rotatebox[origin=c]{90}{Compas}} & Base & 69.89 & 70.58 & 71.65 & 71.46 & 0\\
		& TVAE & 64.42 & 68.07 & 64.33 & 68.33 & 0.0159\\
		& CTGAN & 58.14 & 60.21 & 59.5 & 58.12 & 0.0373 \\
		& WGANWC & 65.34 & 66.5 & 65.06 & \textbf{68.46} & \textbf{0.0095}\\
		& WGANGP & 64.12 & 66.91 &65.20& 68.34 & 0.0179\\
		\bottomrule
	\end{tabular}%
\end{table}%

\subsection{Synthetic Data Utility}

We evaluate machine learning efficacy of synthetic data generated by four generative models, CTGAN~\cite{xu2019modeling}, TVAE~\cite{xu2019modeling}, WGAN-GP and WGAN-WC vary four binary classifiers: \verb|DecisionTreeClassifier|, \verb|MLPClassifier|, \verb|AdaBoostClassifier|, and \verb|LogisticRegression| (Standard scikit-learn machine learning library, see middle columns in Table~\ref{tab:ML}). We also compare ECDFs using the average of all attribute-wise Wasserstein distance $\mathbb{E}_i (l_{1})$ (see the last column in Table~\ref{tab:ML}). All numerical features are min-max scaled to $(0,1)$ and categorical features are one-hot encoded before feeding into the classifier. For ``base'', we trained on the sensitive dataset $D_t$ that is used for data synthesis and test on the real test set $D_s$ (see Table~\ref{tab:dataset}). For synthetic data, we trained on a synthetic dataset $S$ of the same size as the sensitive dataset and test on the same real test set $D_s$. To implement CTGAN and TVAE, we directly feed our pre-processed data into the module \verb|CTGANSynthesizer| and \verb|TVAESynthesizer| of the \verb|SDGym|~\cite{SDGym} (published code for ~\cite{xu2019modeling}). 

According to Table~\ref{tab:ML}, the synthetic dataset generated by WGAN-GP, WGAN-WC, CTGAN and TVAE can greatly restore the prediction ability of the model trained on original dataset. TVAE is least ideal than the others in the Adult dataset. CTGAN is least ideal than the others in the Compas dataset. We will use these learned generative models to perform \emph{TableGAN-MCA} experiments later. 

\begin{figure}[!t]
	\centering  
	\subfigure[Adult]{\includegraphics[width=\columnwidth]{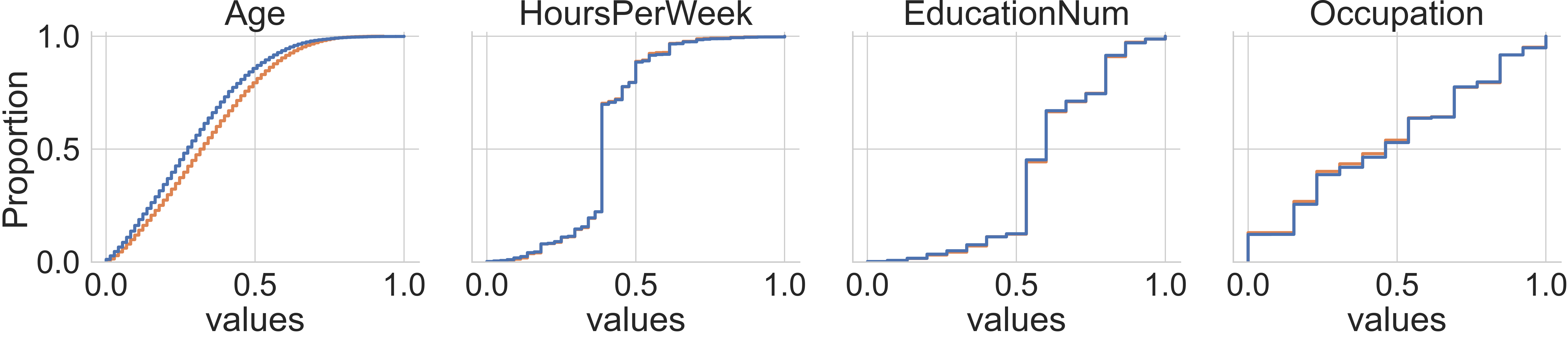}}
	\vspace{-1mm}
	\subfigure[Compas]{\includegraphics[width=\columnwidth]{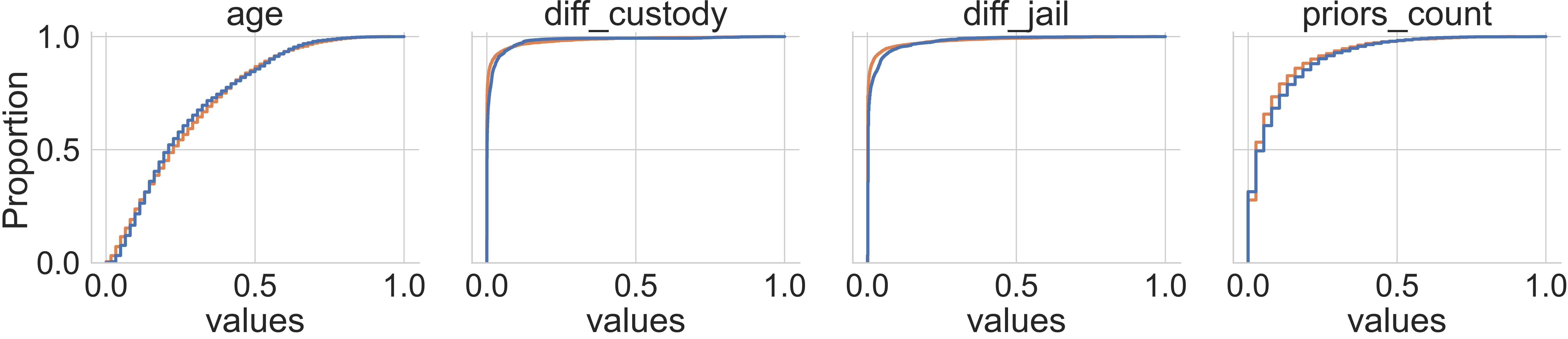}}
	\vspace{-1mm}
	\subfigure[Lawschool]{\includegraphics[width=\columnwidth]{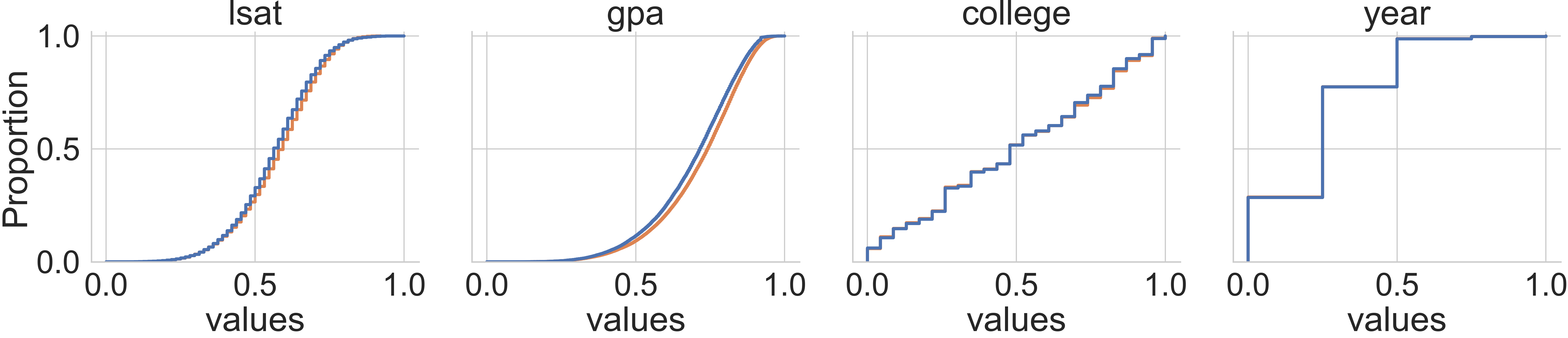}}
	\vspace{-2mm}
	\caption{The Empirical Cumulative Distribution of each attribute in the Adult, Compas and Lawschool datasets (Orange line for real and blue line for synthetic).}
	\label{fig:CDF}	
\end{figure}

For marginal fitness, we depict an additional ECDF comparison between real and synthetic Adult, Lawschool and Compas datasets generated by WGAN-GP in Fig.~\ref{fig:CDF}. In our experiments, we depict ECDFs of continuous variables (i.e., age, isat) and more complex categorical variables (i.e., hours per week, priors count) since they are more difficult to fit. 
As can be seen in Fig.~\ref{fig:CDF}, the marginals of the synthetic dataset are almost indistinguishable from the original one, thus supporting any statistical queries. 

 \vspace{-3mm}
\subsection{Attack Performance}
\subsubsection{Performance Evaluation on \emph{TableGAN-MCA}}
\label{attack_eval}
In this section, we evaluate \emph{TableGAN-MCA} of Alg.~\ref{alg:TableGAN-MCA} on the Adult, Lawschool and Compas datasets. The training and inference data statistics of \emph{TableGAN-MCA} are presented in Table~\ref{tab:statistics}, where positive percentage implies the membership collisions proportion. Both target models and shadow models are WGAN-GP. The attack model is trained on the shadow dataset $\widetilde{S}$ and tested on the synthetic dataset $S$. 

\begin{table}[!t]
  \centering
   \caption{Training and inference statistics for the Adult, Compas and Lawschool datasets in \emph{TableGAN-MCA}.}
   \vspace{-1mm}
  \label{tab:statistics}%
    \begin{tabular}{ccccc}
    \toprule
         & Adult  & Lawsch & Compas\\
    \midrule
    $|\widetilde{S}|$ (Train) & $31655$  &$43011$& $3694$ \\
    $|S|$ (Inference) & $31655$  &$43011$& $3694$ \\
    \hline
    $\operatorname{Pr}_{\widetilde{S}}[y_i=1]$ & $15.99\%$ & $22.68\%$ &$40.49\%$ \\
    $\operatorname{Pr}_{S}[y_i=1]$ & $16.90\%$ & $23.89\%$ &$34.00\%$ \\ 
    \bottomrule
    \end{tabular}%
\end{table}%

\begin{figure}[!t]
	\centering  
	\includegraphics[width=0.8\columnwidth]{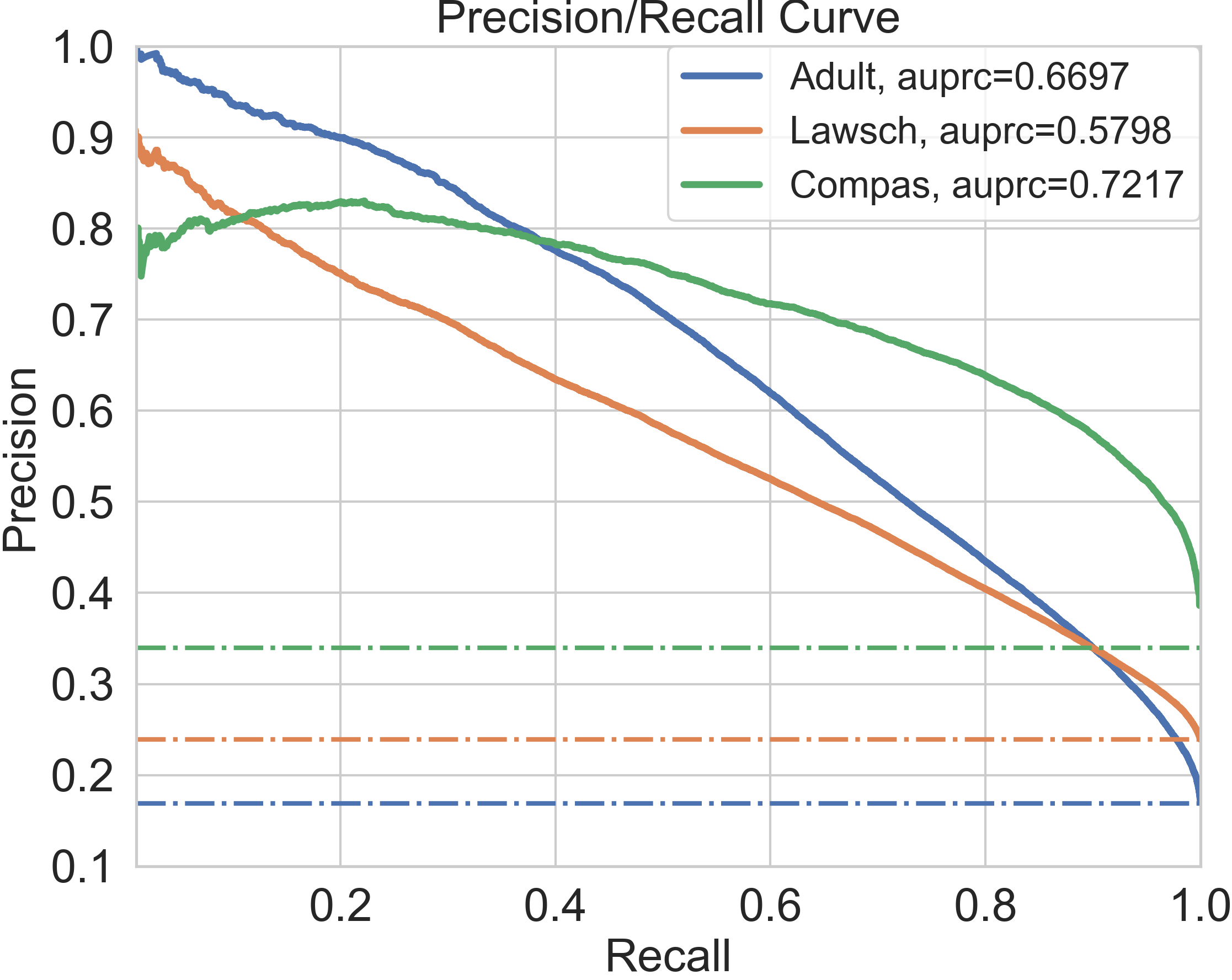}
	\caption{Attack effect of \emph{TableGAN-MCA}. The dash-dot lines imply random guess baselines ($0.1690$, $0.2389$, $0.3400$ for Adult, Lawschool and Compas datasets, respectively.)}
	\label{fig:PRCurve}
  \vspace{-2mm}
\end{figure}

\textbf{\emph{TableGAN-MCA} provides a promising attack against the GAN-synthesized tables.} We report the PR-curve of the attack model in Fig.~\ref{fig:PRCurve} when $N_s =1$, i.e., $|S|=|D_t|$. In Fig.~\ref{fig:PRCurve}, PR-curve reflects the trade-off between precision and recall for different probability thresholds ${\rm{T}}$. Particularly, after providing the inference data (the released GAN-synthesized tables) to the \emph{TableGAN-MCA} attack model, we receive a set of probabilities for each record of the test data that predicts whether a record is a member. 

As illustrated in Fig.~\ref{fig:PRCurve}, we find that by setting a suitable threshold ${\rm{T}}$, the adversary can expose approximate $30\%$ colliding members with confidence over $83.91\%$, $69.40\%$ and $81.24\%$ for the Adult, Lawschool and Compas datasets, respectively. This means that the adversary significantly increases its ability to assert that these entities are members. Furthermore, when setting confidence to $80\%$, we have $36.1\%$, $12.7\%$, $36.5\%$ positive percentages being exposed, which correspond to $1931$, $1304$ and $458$ individual's sensitive entries in the Adult, Lawschool and Compas datasets, respectively. According to Fig.~\ref{fig:PRCurve}, we list \emph{TableGAN-MCA}'s recovery rates (Eq.\eqref{ep:recovery}) with different precision configurations in Table~\ref{recovery_rate}. 

\begin{table}[!t]
  \centering
  \caption{\emph{TableGAN-MCA}'s recovery rate $\rho_{\mathcal{A}}$. ($|R_{\mathcal{A}}|$: $\#$ of recovered data points under attack algorithm $\mathcal{A}$)}
  \vspace{-1mm}
  \label{recovery_rate}%
    \begin{tabular}{cccccc}
    \toprule
    Datasets & $\rho_{\mathcal{A}}$($\%$) & $|R_{\mathcal{A}}|$  & $|D_t|$ & Precision & Recall\\
    \midrule
   \multirow{2}{*}{Adult} & 
      $3.04$  & $962$ & $31655$ &  $0.9$ & $0.16$ \\
     & $6.10$  & $1931$ & $31655$ & $0.8$  & $0.36$\\ 
     \hline
  \multirow{2}{*}{Lawsch} & $3.03$  & $1305$ & $43011$ & $0.8$  & $0.13$\\
   & $4.66$  & $2003$ & $43011$ & $0.75$ & $0.18$\\
   \hline
  \multirow{2}{*}{Compas} & $12.41$  & $458$ & $3694$ & $0.8$ & $0.37$\\
   & $17.17$  & $634$ &$3694$ & $0.75$ & $0.43$\\
    \bottomrule
    \end{tabular}%
\end{table}%

\begin{figure*}
\centering  
\subfigure[Adult PR-curve]{\includegraphics[width=0.24\textwidth]{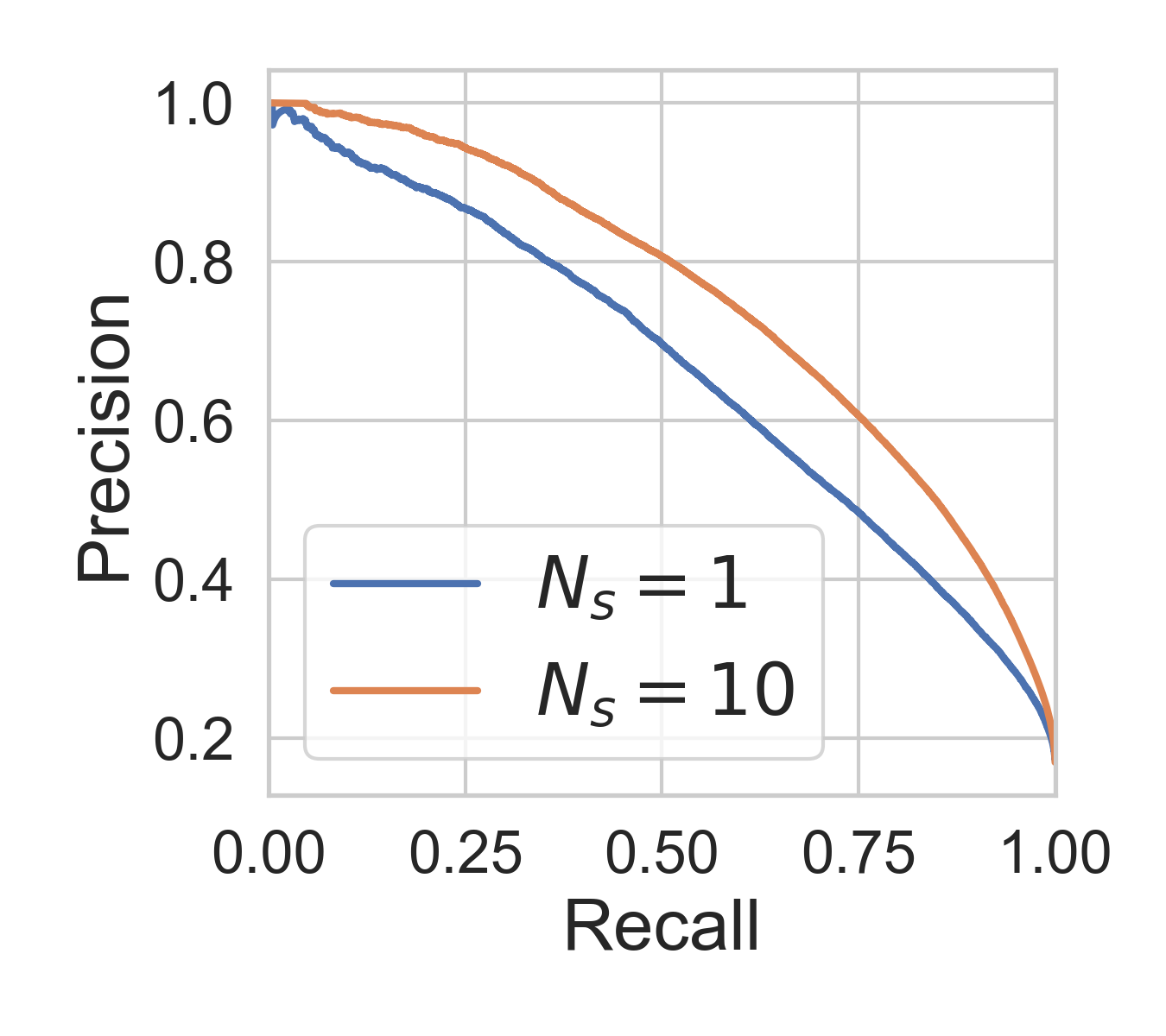}}
\subfigure[Lawsch PR-curve]{\includegraphics[width=0.24\textwidth]{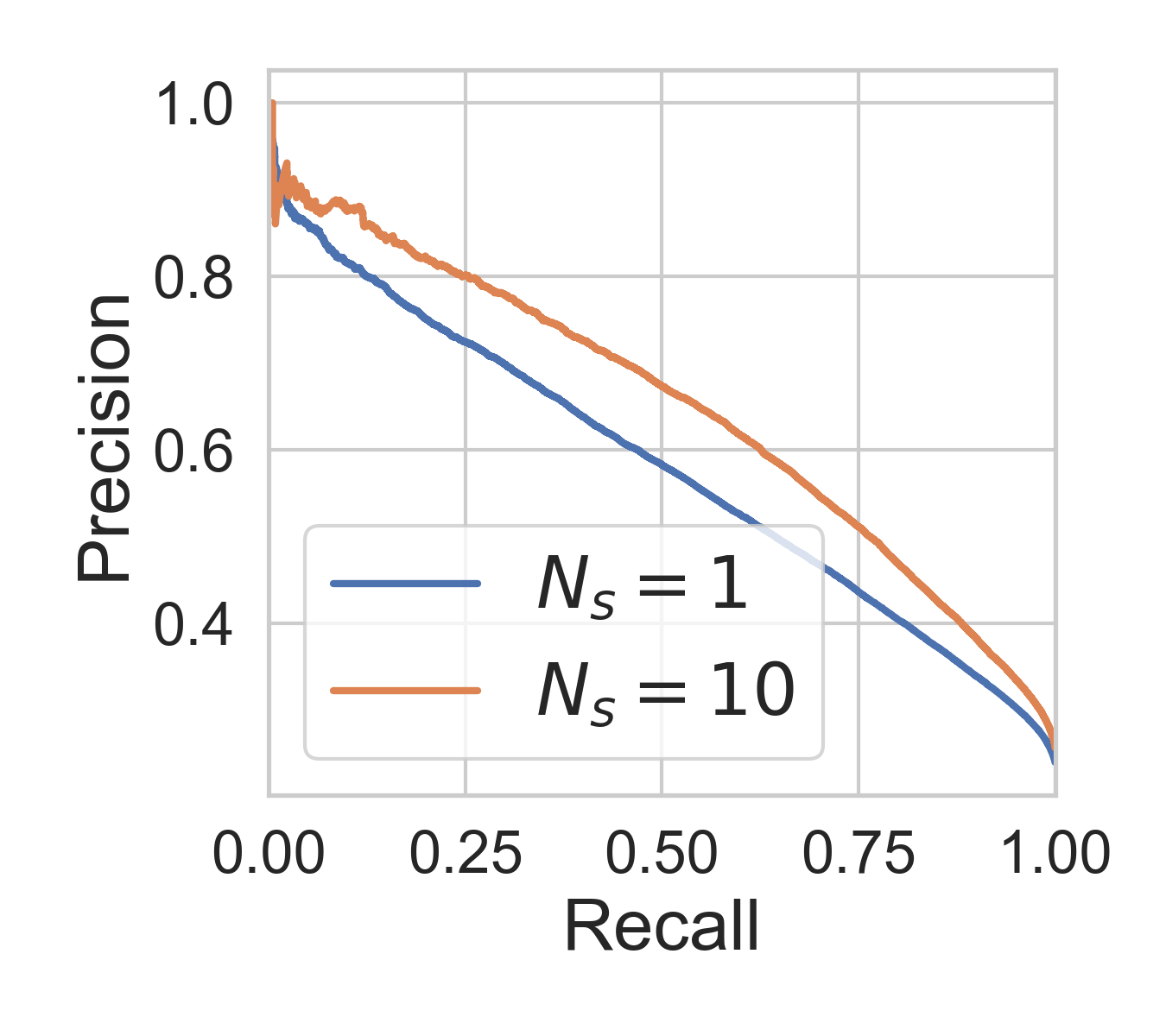}}
\subfigure[Compas PR-curve]{\includegraphics[width=0.24\textwidth]{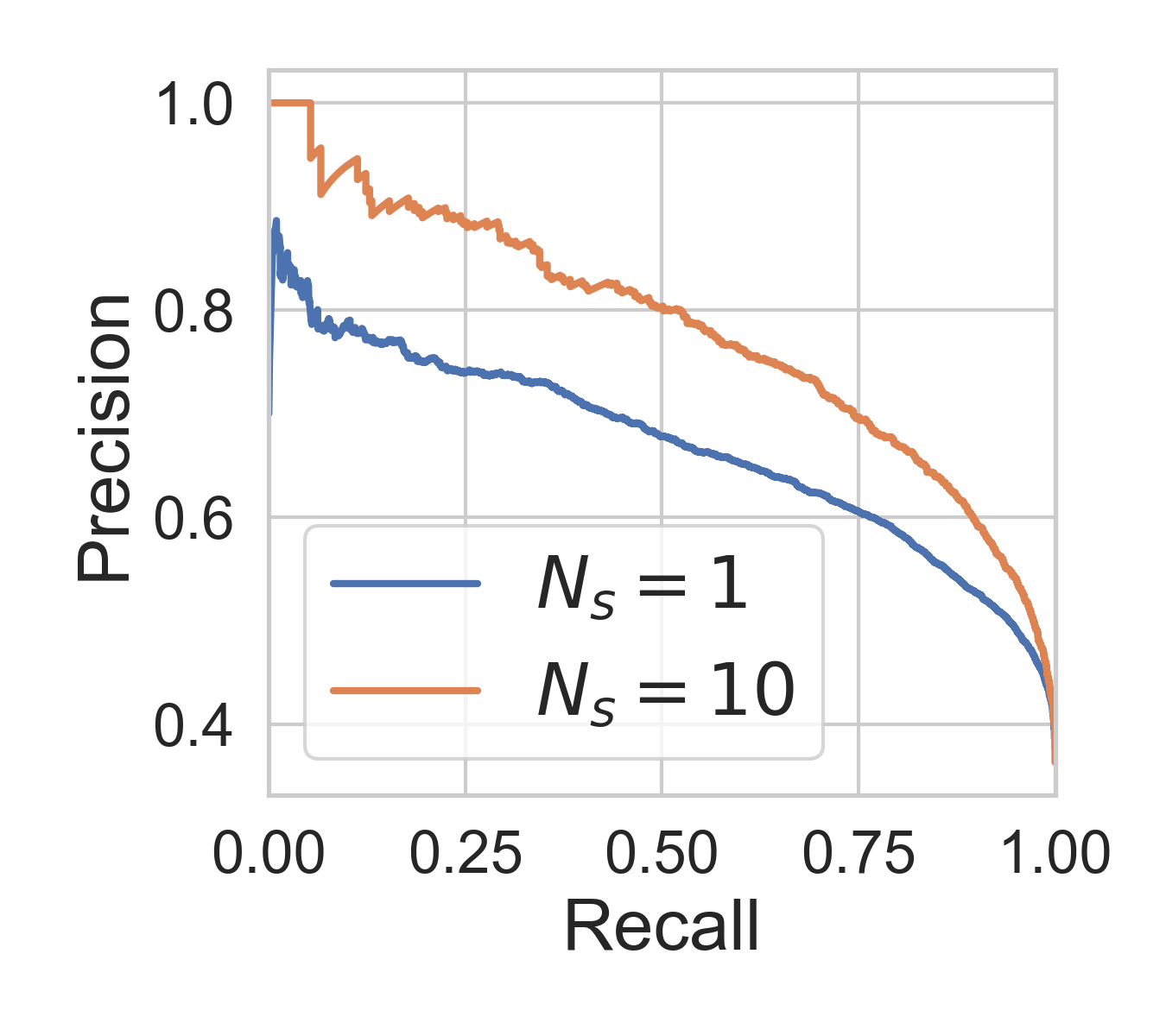}}
\subfigure[AUPRC]{\includegraphics[width=0.24\textwidth]{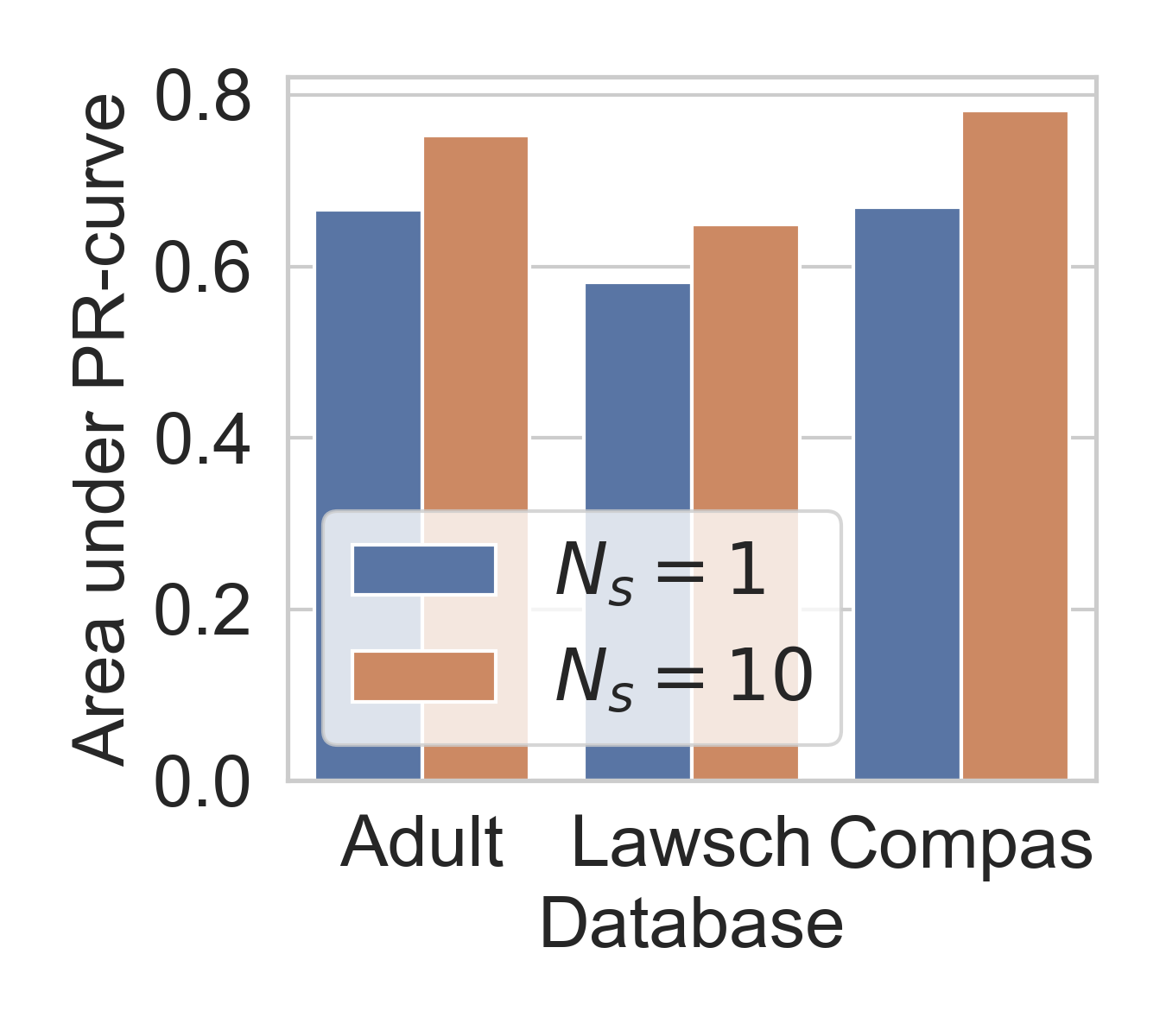}}
\vspace{-2mm}
\caption{Attack performance comparisons between one synthetic copy ($N_s=1$) and ten synthetic copies ($N_s=10$) in \emph{TableGAN-MCA}. (a), (b), (c): PR-curve comparison for three datasets; (d): AUPRC comparisons for three benchmarks.}
\label{fig:boost_attack}
\end{figure*}

\textbf{Adversary's knowledge enhances the attack performance of \emph{TableGAN-MCA}.}  Fig.~\ref{fig:boost_attack} reports the PR-curve and AUPRC of \emph{TableGAN-MCA} when $N_{s} = 10$. That is, the adversary has multiple copies of the released synthetic data. In particular, when $N_s =10$, the adversary trains $10$ independent shadow GAN for each one of $S_i$ and finally obtains a shadow dataset $N_s^2$ times the size of $|D_t|$. In Fig.~\ref{fig:boost_attack}, the performance of \emph{TableGAN-MCA} is greatly improved by increasing the number of synthetic copies $N_s$. We also show the PR-curve comparison for the three datasets in Fig.~\ref{fig:boost_attack}. That is, given $30\%$ recall, 10 copies boost the precision from $86.81\%$ ($N_s=1$) to $90.70\%$ ($N_s=10$). Given $90\%$ precision, the recall is boosted from $22.25\%$ ($N_s=1$) to $33.01\%$ ($N_s=10$). We conclude that more copies of the synthetic data allow the adversary to make better approximation to generated distribution $\mathbf{P}_g$, and thus generate more informative labeled shadow samples to train the attack model. 

\textbf{TableGAN-MCA achieves commendable attack performance even with fewer synthetic queries.} When $N_s=0.25$ (an adversary queries the target Generator $0.25*|D_t|$ times),  \emph{TableGAN-MCA} achieves $0.6674$ AUPRC ($N_s=1$ is 0.6697), and recovers 1,409 data points ($N_s=1$ is 1565) under $75\%$ precision in the Adult dataset. More details are shown in Appendix~\ref{LSQ}.

\textbf{The generation quality of the target/victim model positively impacts attack performance.} Fig.~\ref{fig:auprc_targets} depicts \emph{TableGAN-MCA}'s performance on four different target/victim models: WGAN-GP, WGAN-WC, CTGAN and TVAE. The shadow model in use is exactly the same as target models. Combining the results of Fig.~\ref{fig:auprc_targets} and Table~\ref{tab:ML}, we conclude that target/victim generators with high generation quality often attain high attack performance. For instance, TVAE with the lowest prediction accuracy score in prediction accuracy (see Table~\ref{tab:ML}) also achieves unsatisfactory performance in \emph{TableGAN-MCA} on the adult dataset. This echoes what CTGAN performs in the Compas dataset. Additionally, we observe that attack performance of TVAE is more sensitive to its generation quality.

\begin{figure}[!t]
	\centering  
	\subfigure[Different Target/Victim Models]
{\includegraphics[width=0.48\columnwidth]{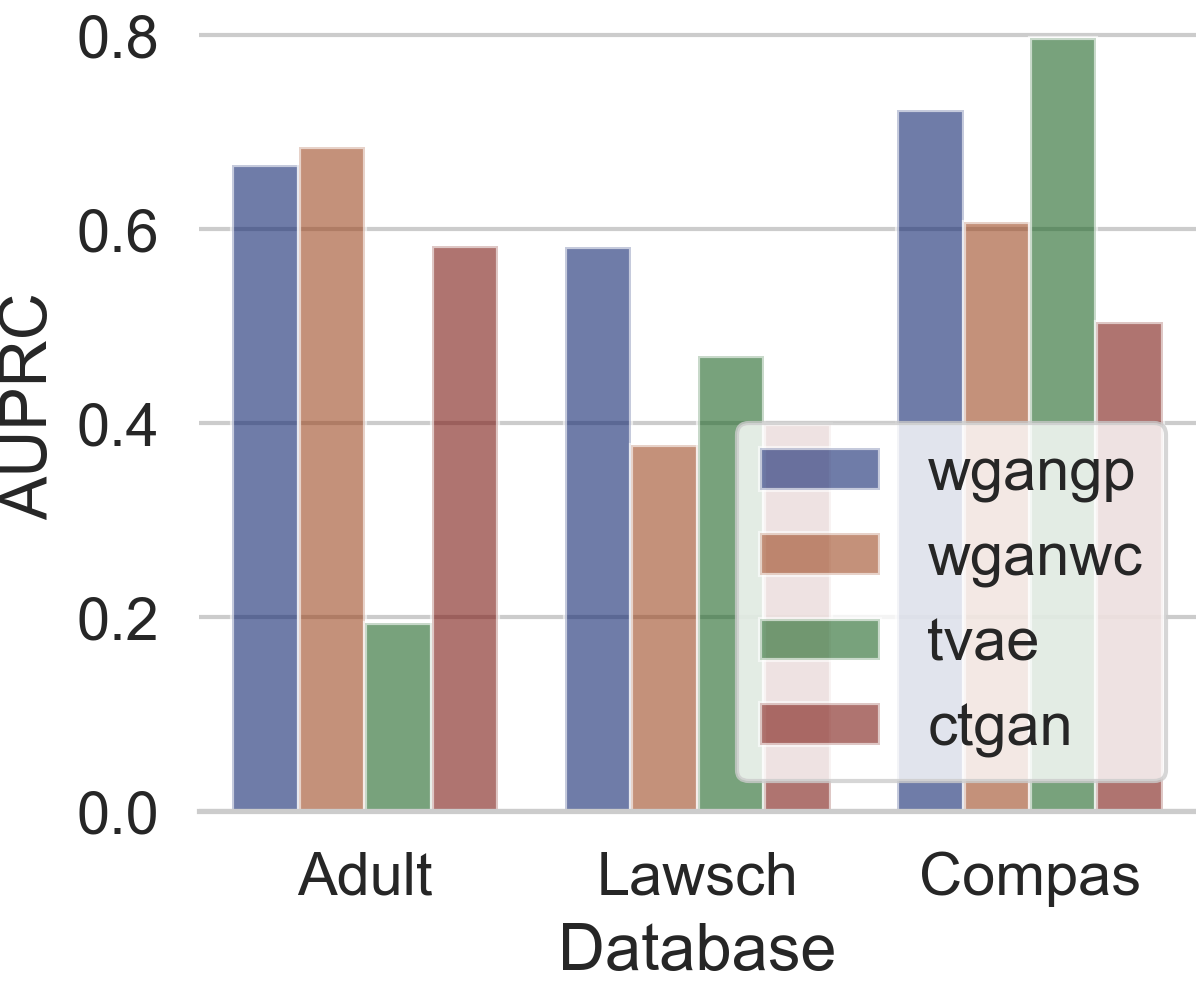}
\label{fig:auprc_targets}
 }
	\subfigure[Different Shadow Models]
{\includegraphics[width=0.48\columnwidth]{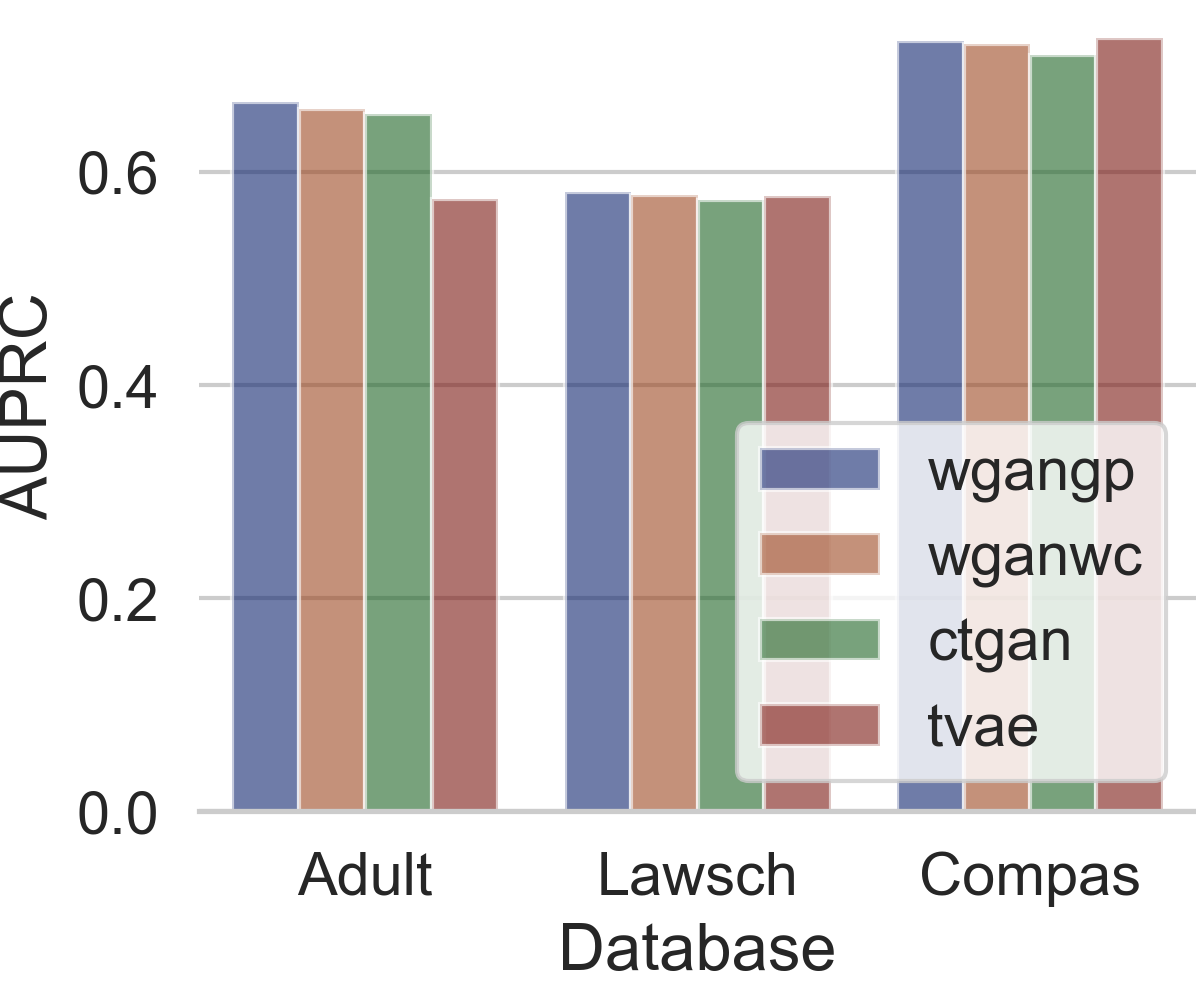}
\label{fig:auprc_shadows}
}
	\vspace{-1mm}
	\caption{Attack effect of \emph{TableGAN-MCA} under different target/victim-shadow model settings. In (a), shadow models are the same as target/victim models.}
	\label{fig:auprc_targets}
\end{figure}

\textbf{The type of shadow models has limited impact on attack performance}. Note that the adversary may have no knowledge about the structures and parameters of target/victim generative models. Fig.~\ref{fig:auprc_shadows} compares the attack performance by using four different shadow models (WGAN-GP, WGAN-WC, CTGAN and TVAE) to attack target WGAN-GP. As can be seen, various shadow model attacks (``wganwc'', ``ctgan'', ``tvae'') work as well as the identical shadow model attack (``wgangp'').  TVAE shadow models perform worst, in large part due to its poor learning ability in the Adult dataset.

\textbf{The success of \emph{TableGAN-MCA} is mainly due to the observed} collision, the membership collisions indicator and the shadow model in use. In particular, the collision between synthetic data and training set provides the opportunity for recovering training data. The membership collisions indicator, which captures the statistical patterns behind colliding members, guarantees more accurate and informative features for training the attack model. The shadow model in use provides enough labeled data to train the attack model so as to learn from the statistical patterns of the colliding members.

\textbf{Attack scalability.} The key to the success of the \emph{TableGAN-MCA} is the possibility of collisions between raw training datasets and synthetic datasets. For real-world tabular data, attributes usually have a finite domain range. Hence, the dataset dimension indicates its overall domain range. Namely, low-dimensional tables are more likely to incur sample collisions when the generator creates those synthetic tables. Therefore, \emph{TableGAN-MCA} discovers additional privacy risks -- membership disclosure via collision attacks -- for low-dimensional data. \emph{TableGAN-MCA} potentially fits high-dimensional data if adversaries reduce the data granularity by generalizing attributes. The \emph{TableGAN-MCA} works since the synthetic data would have a higher chance to collide with the training datasets. In our experiments, \emph{TableGAN-MCA} achieves 0.871 AUPRC by bucketizing the "age" attribute in synthetic Adult datasets into 10 bins (no-bucketization baseline: 0.668).

\vspace{-1mm}
\subsubsection{Comparisons between \emph{TableGAN-MCA} and existing MIAs}
Firstly, \textbf{\emph{TableGAN-MCA} recovers member data points from the GAN-synthesized tables previously  assumed to be resilient to table-GAN~\cite{park2018data}.} 
We evaluate the performance of table-GAN against the same WGAN-GP that used in \emph{TableGAN-MCA} evaluation. Notice that we test their MIAs directly on the target discriminator instead of the shadow discriminator due to the fact that if the target discriminator fails, the shadow model will perform even worse. We report the accuracy of membership prediction (member/non-member) of table-GAN, which are $50.17\%$, $50.80\%$ and $50.67\%$ for Adult, Lawschool and Compas datasets, respectively ($50\%$ is the baseline of random guess). Taken altogether the experiment results in Fig.~\ref{fig:PRCurve}, we conclude that GAN APIs with a black-box access assumed to be resilient to table-GAN (targeting on a discriminative model)~\cite{park2018data} may still disclose partial sensitive training information under \emph{TableGAN-MCA}. 

Secondly, \textbf{the MIAs proposed in GAN-Leaks and LOGAN cannot disclose membership collisions.} Note that the existing MIAs against GANs may work in the MCA scenario. Thus, we perform additional experiments to infer membership collisions of each synthetic data point using their methods. In particular, we evaluate LOGAN (black-box attack with no auxiliary knowledge) and GAN-Leaks (full black-box generator attacks) under threat model (1) (given one copy synthetic data, see details in Section~\ref{sec:threat_model}) and report the result in Table~\ref{AP_compare}. MC and table-GAN are not included in this experiment. The reason is two-fold. First, the distance function of MC is not directly applicable to non-image datasets. Second, table-GAN requires predicted probability vectors of the target discriminator, which is not permitted in our threat model. Note that the synthetic dataset $S$ has imbalanced membership collisions labels (Row 1 in Table~\ref{AP_compare}) that are different from Shokri's shadow model MIA~\cite{shokri2017membership} (random observation with $50\%$ real members) since the number of colliding data points (members) is usually unequal to non-colliding ones (non-members). 

We observed that the results in GAN-Leaks are close to the random guess baseline. This is due to the reconstruction loss $L(x,x*)=0$ for all synthetic data regardless of membership collisions (the optimal reconstruction of a synthetic data x is itself). Furthermore, LOGAN did not show convincing inference results since it never learns the intersection between the synthetic data and the private training data. In comparison, \emph{TableGAN-MCA} learns such an intersection (by which we recover partial training data) through the intersections of the published synthetic data (by mimicking the private training data) and shadow (synthetic) data (by mimicking the original synthetic data).

In summary, the MIA classifiers that identify membership fail to identify those membership collisions since the decision boundaries of our attack classifier is different from those of MIAs against GANs.

\begin{table}[!t]
  \centering
  \caption{The attack AUPRC comparison (mean $\pm$ SD). Base implies random guess baseline. We use WGAN-GP as target/victim as well as shadow models.}
  \vspace{-1mm}
  \label{AP_compare}%
    \begin{tabular}{cccc}
    \toprule
    & Adult & Lawschool  & Compas \\
    \midrule
   Base & $0.1690 \pm 0.0038$  & $0.2389 \pm 0.0067$ &  $0.3400 \pm 0.0233$  \\ 
   LOGAN  & $0.2237 \pm 0.0194$  & $0.2512 \pm 0.0172$ &  $0.3154 \pm 0.0343$ \\
  GAN-Leaks  & $0.1667 \pm 0.0063 $  & $0.2514 \pm 0.0061 $ & $0.3256 \pm 0.0301$  \\
  Proposed   & $0.6681 \pm 0.0348 $  &  $0.5805 \pm 0.0144 $& $0.7228 \pm 0.0556$  \\
    \bottomrule
    \end{tabular}%
\end{table}%

\section{TableGAN-MCA Analysis}
In this section, we discuss the factors that may impact the attack performance of \emph{TableGAN-MCA} from the following aspects, such as GAN training set size, GAN training epochs and GAN training data frequencies. We choose WGAN-GP as targets as well as shadow model for its superior modeling quality and stability in \emph{TableGAN-MCA} experiments. 

\subsection{GAN Training Set Size}
\begin{figure*}[!t]
\centering  
\subfigure[Adult]{\includegraphics[width=0.32\textwidth]{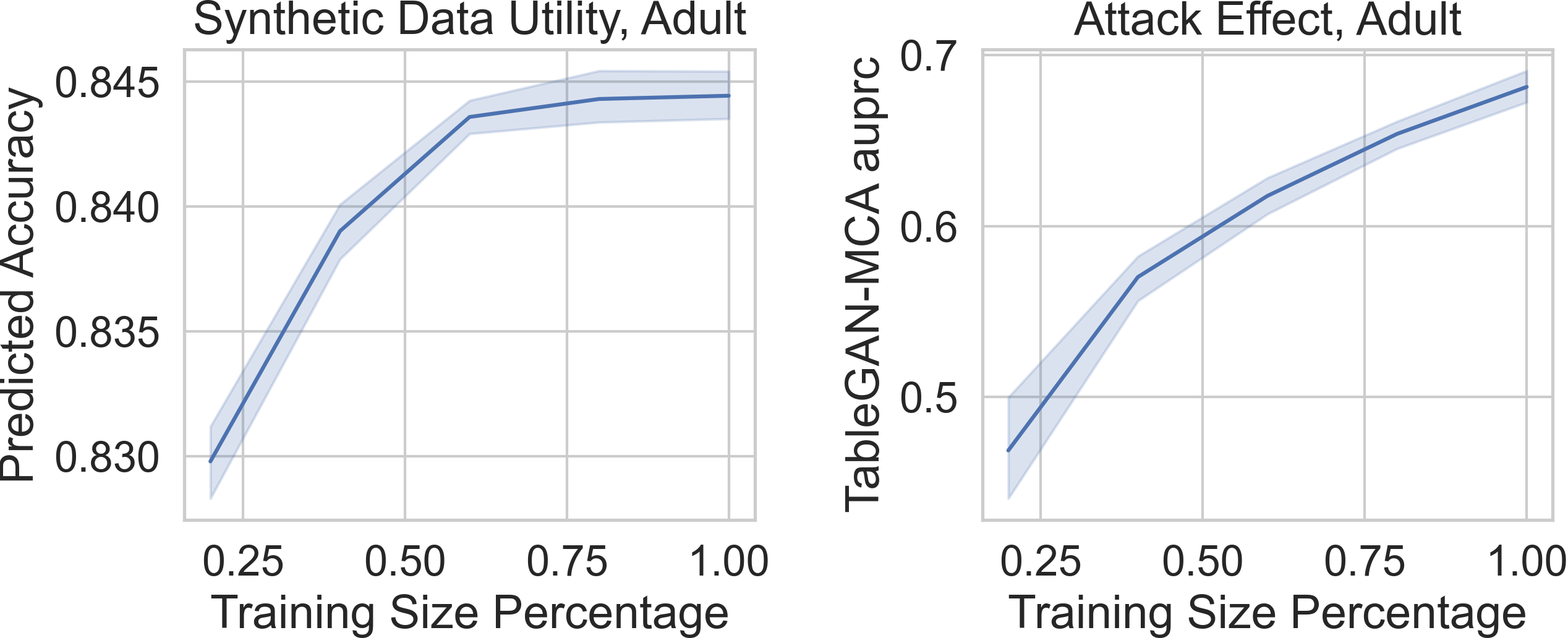}}
\subfigure[Lawschool]{\includegraphics[width=0.32\textwidth]{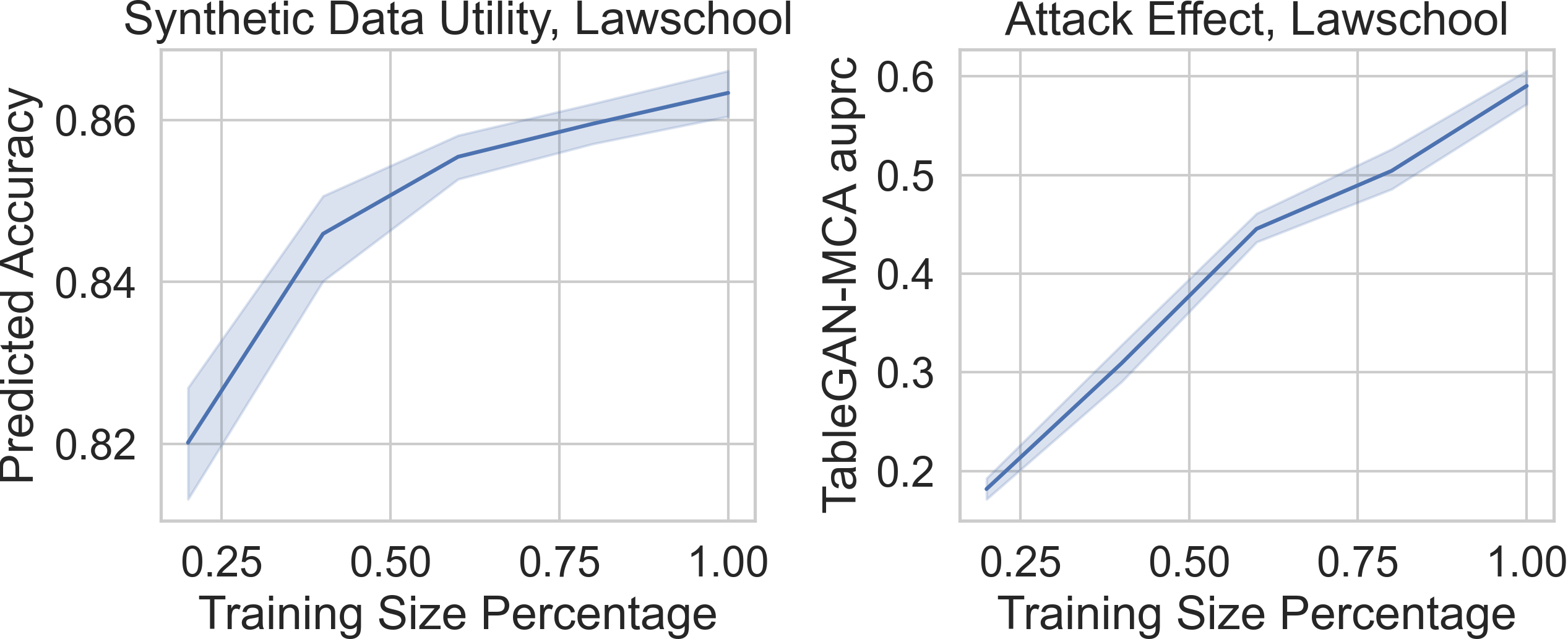}}
\subfigure[Compas]{\includegraphics[width=0.32\textwidth]{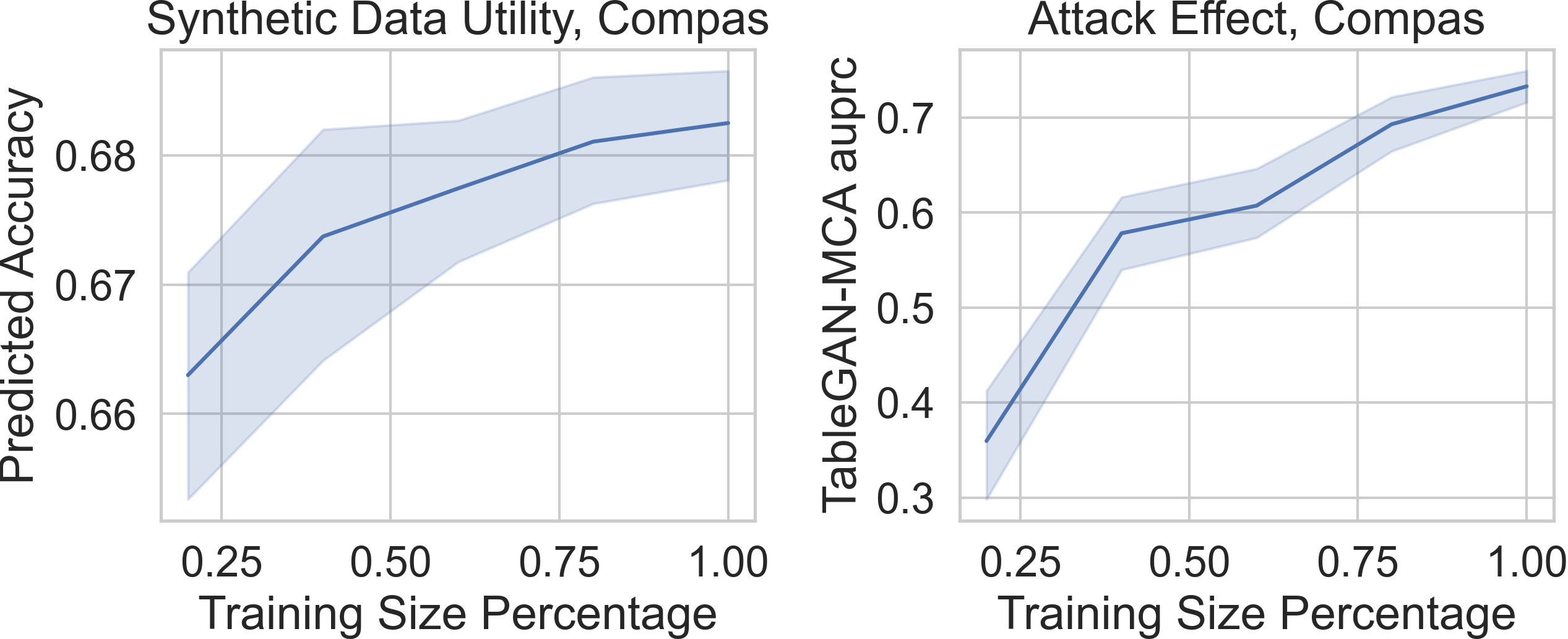}}
\vspace{-3mm}
\caption{The impact of GAN training data size on synthetic data utility (left) and \emph{TableGAN-MCA} effect (right). The x-axis indicates the amount of GAN training data}
\vspace{-3mm}
\label{fig:trainingSize}
\end{figure*}

\textbf{The size of the training dataset for a GAN model positively impacts the attack performance.}
Fig.~\ref{fig:trainingSize} depicts the positive impact of training dataset size on prediction accuracy and AUPRC of \emph{TableGAN-MCA}, where $1.0$ in x-axis indicates the full size of a given dataset, $N_{s} = 1$. Especially, when the size of the training dataset is less than $0.5$ of the full dataset, increasing the size has a significant impact on the attack performance. The intuition behind the experimental results is two-fold. First, less training data decrease the number of colliding members (positives) in fixed amount of synthetic datasets thus decreases the attack effect. Second, GAN learns a less accurate data distribution if trained on a smaller dataset. Synthetic data generated by such a distribution contain less information than the original training data hence hard for the adversary to learn the statistical patterns of the members/non-members. Note that our results do not conflict with~\cite{hayes_logan_2019,chen_gan-leaks_2020,pmlr-v130-lin21b} since we use different measurements (PR space vs ROC space) that focus on different domains~\cite{davis_relationship_2006}. Additionally, our attack target (test data) is also different. We aim to recover the colliding member data from the released synthetic dataset whereas they aim to infer the membership of a random target data point, and thus we learn different decision boundaries.

\subsection{GAN Training Epochs}
\begin{figure*}[!t]
\centering 
\subfigure[Adult]{
    \includegraphics[width=0.32\textwidth]{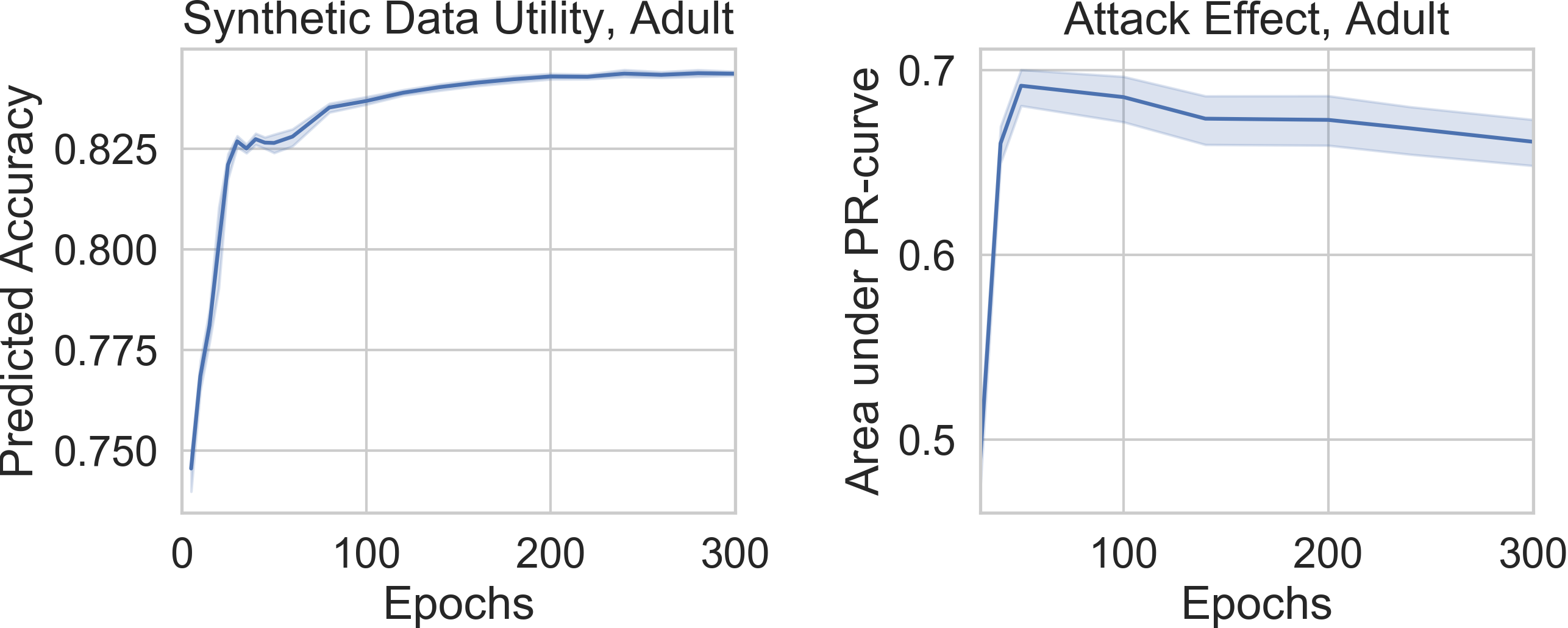}%
}
\subfigure[Lawschool]{
    \includegraphics[width=0.32\textwidth]{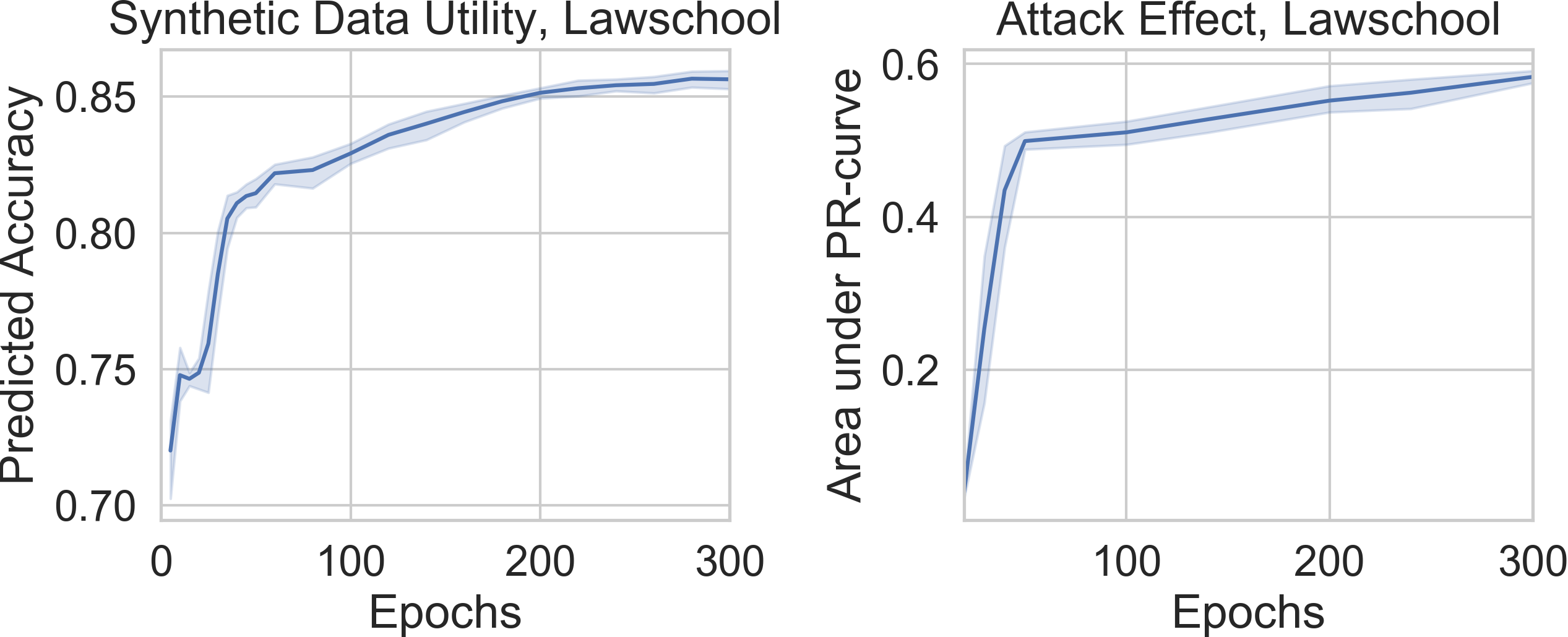}%
}
\subfigure[Compas]{
	\includegraphics[width=0.32\textwidth]{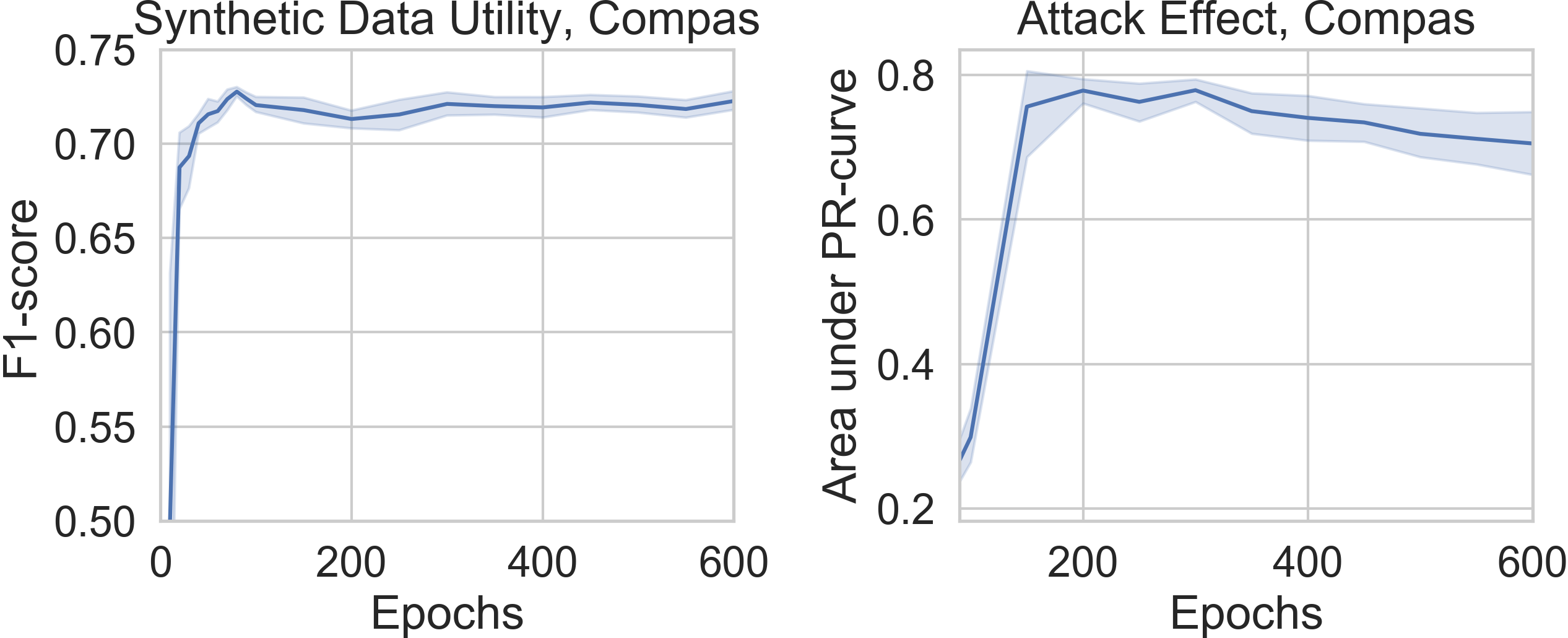}
}
\vspace{-3mm}
\caption{The impact of GAN training epochs on Synthetic data utility (left) and \emph{TableGAN-MCA} effect (right).}
\label{fig:attack_epoch}
\end{figure*}

\textbf{Epochs impact the attack performance of \emph{TableGAN-MCA} by impacting the knowledge learned by GAN models.} We study the attack performance on different training stages by setting different epochs in Fig.~\ref{fig:attack_epoch}, where we report the attack prediction accuracy and attack AUPRC.

As seen from Fig.~\ref{fig:attack_epoch}, we find that the membership leakage starts at the very beginning of the training epoch, even before the GAN reaches the Nash equilibrium. Interestingly, in Adult and Compas, the attack effect seems to slightly decrease when we set a larger epochs for training GAN models. Since \emph{TableGAN-MCA} tends to recover the data with high appearance frequency (recall Fig.~\ref{fig:observation}), we conclude that with increasing epochs, GAN models learn more about the training data distribution; hence, the released synthetic data contain more information, which enhances the attack performance. However, once the GAN models learn the details of the data distribution, such details about the distribution would dilute the frequency of those data supposed to have high frequency. The attack performance of \emph{TableGAN-MCA} is then potentially dropped.

\subsection{Training Data Frequencies} 

\begin{figure}
 \centering  
 \includegraphics[width=0.8\columnwidth]{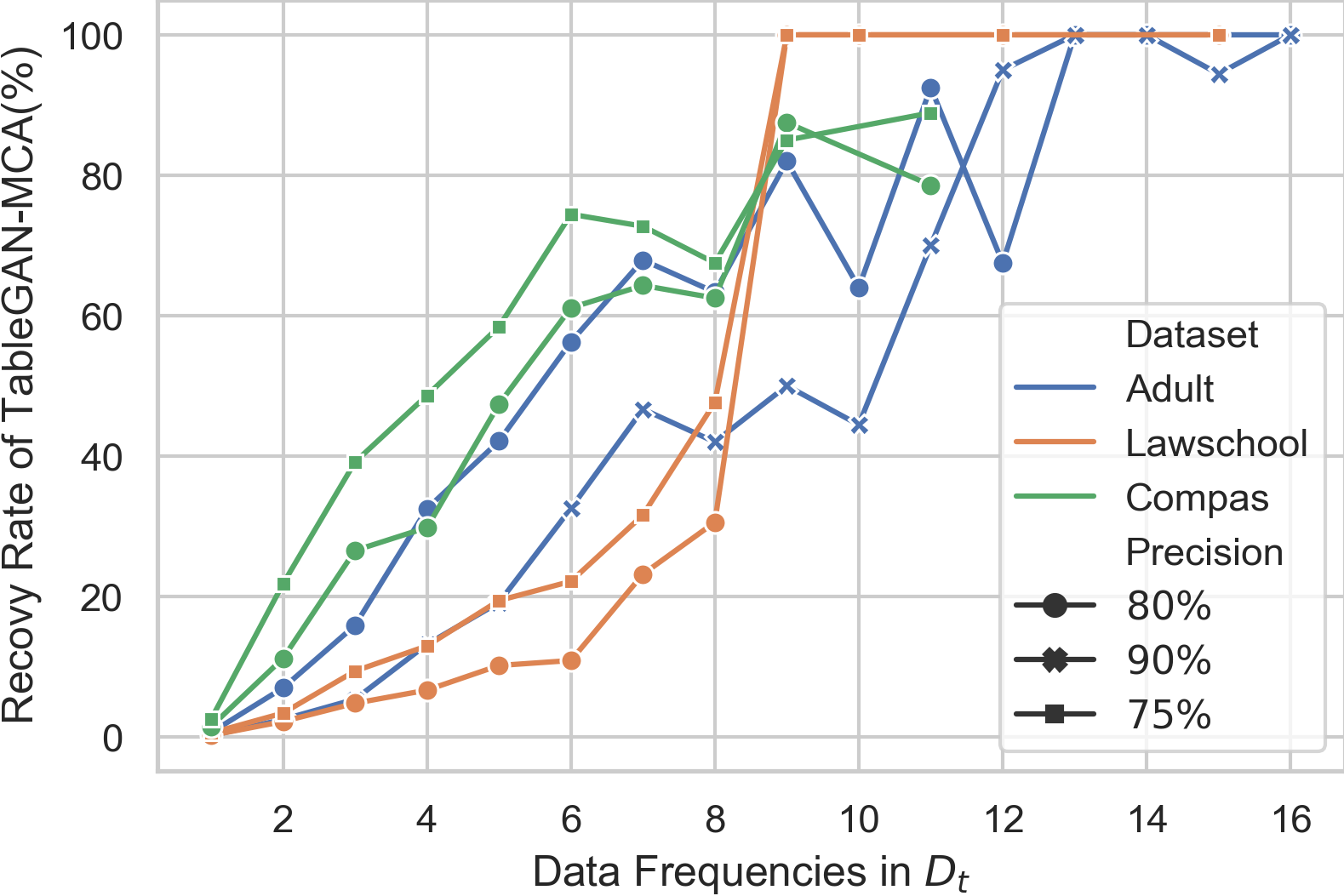}
 \caption{The impact of training data frequencies on \emph{TableGAN-MCA} effectiveness. The attack precision is set to be one of $\{75\%, 80\%, 90\%\}$.}
 \label{fig:exposed_members}

 \end{figure}

\textbf{Training data frequencies are positively correlated with training data recovery probabilities by \emph{TableGAN-MCA}.} We first compute the recovery possibility and appearance frequency for each data point. We then plot the recovery possibility over the values of data points frequency in Fig.~\ref{fig:exposed_members}. For each dataset, we set two precision-scores of \emph{TableGAN-MCA} and plot the training data frequency-recovery rate curves. 
Overall, highly frequent training data are more susceptible to \emph{TableGAN-MCA}. For instance, when attacking Adult datasets with $80\%$ precision, $41.5\% (784/1892)$ training data with appearance more than three times are recovered by \emph{TableGAN-MCA} whereas only $0.6\% (510/25130)$ of unique training data ($\# \mathbf{x}=1$) are recovered by \emph{TableGAN-MCA}.

For highly frequent training data, GANs inevitably learn and output these common representations frequently; thus it is easy to recover such highly frequent data by \emph{TableGAN-MCA}. The re-identification threats of these data caused by \emph{TableGAN-MCA} are limited since each of them correspond to several individuals and lack of uniqueness.

Unique training data, on the other hand, have more risks for being linked to specific people once recovered by \emph{TableGAN-MCA}. Therefore, it deserves further exploration for the reason of being exposed.

\textbf{Generalization of GAN models may accidentally increase the appearance of some unique data points in the synthetic data, therefore increasing their probability to be recovered by \emph{TableGAN-MCA}.} 
Since the \emph{TableGAN-MCA} is based on data density in modeled distribution $\mathbf{P}_g$, for a recovered unique training data point $\mathbf{x}_i$, we study how \emph{TableGAN-MCA} is impacted by the difference between data density of $\mathbf{x}_i$ in the training distribution~$\mathbf{P}_r$ and that of the modeled distribution $\mathbf{P}_g$. 

\begin{figure*}
\centering  
\subfigure[Adult]{
\includegraphics[width=0.32\textwidth]{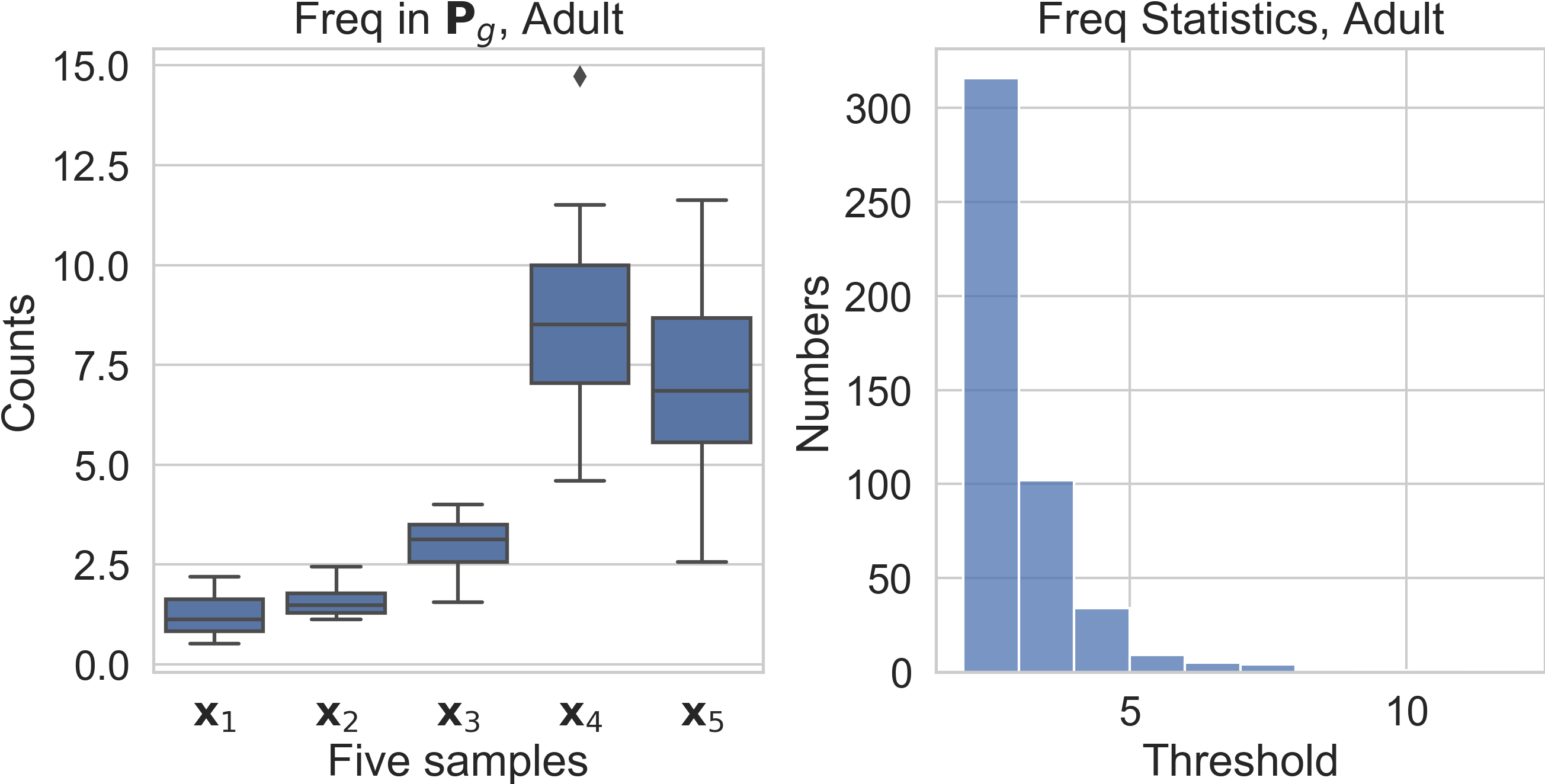}%
\label{impact_adult}
}
\subfigure[Lawschool]{
\includegraphics[width=0.32\textwidth]{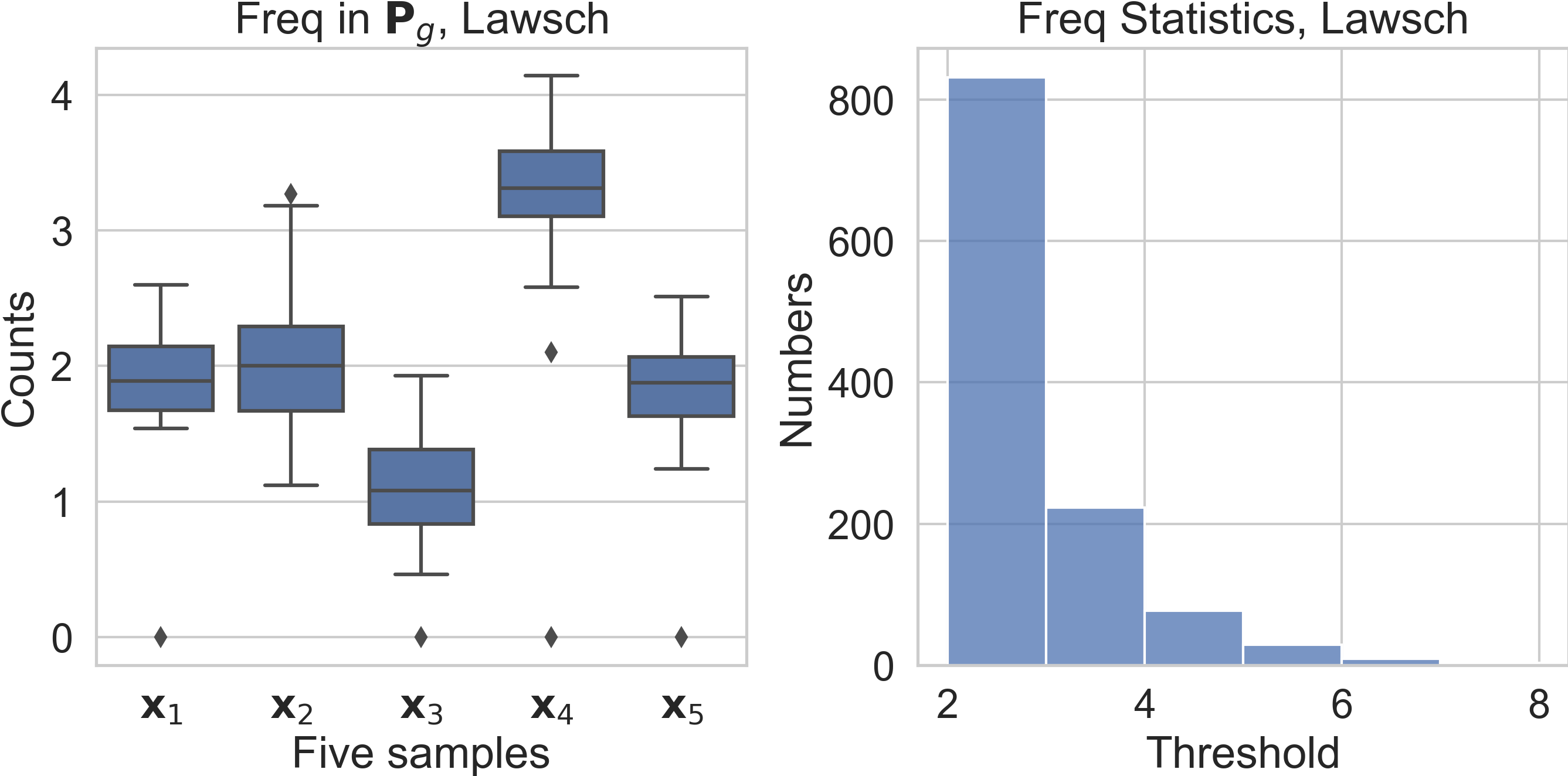}%
\label{impact_lawsch}
}
\subfigure[Compas]{
\includegraphics[width=0.32\textwidth]{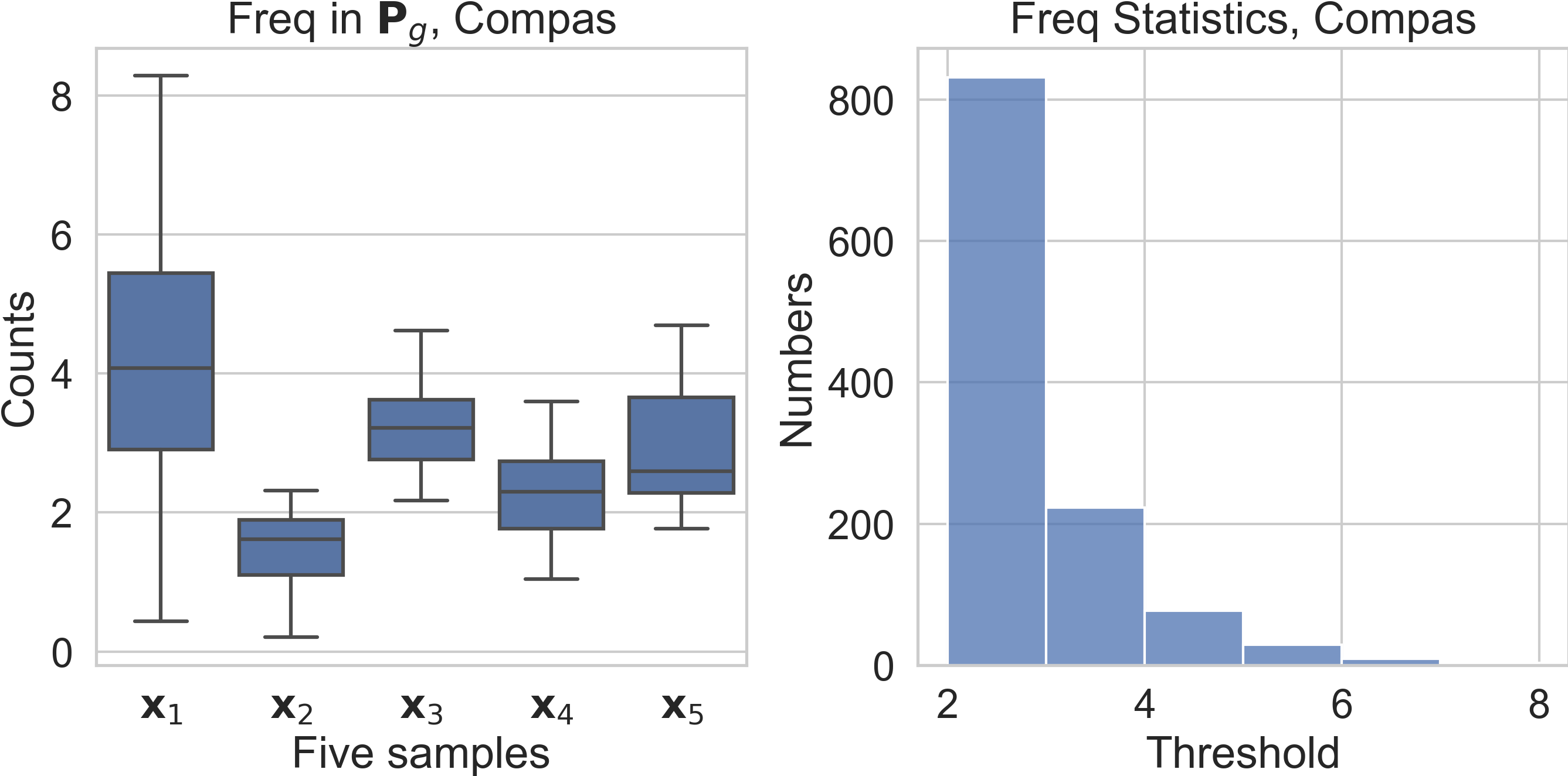}%
\label{impact_compas}
}
\vspace{-3mm}
\caption{Five synthetic samples' count estimation in generated distribution (left) and statistics of data points that increase the frequencies in $\mathbf{P}_g$ across three datasets  (right).}
\label{fig:impact}
\end{figure*}

According to our experiments, we discover that some unique training data ($\forall \mathbf{x}\in \mathbf{P}_r$, $\#\mathbf{x}_i=1$) have unexpected high exposure in modeled distribution $\mathbf{P}_g$. For example, in Fig.~\ref{fig:impact}, we illustrate the average counts (from $100$ synthetic datasets following modeled distribution $\mathbf{P}_g$) of five data points that appear in the training dataset only once. As we can see, these five data points have higher counts than what they have in the training dataset ($= 1$). 
Such an observation indicates that the generator of GAN models unfairly increases the probability of exposure of some data points under \emph{TableGAN-MCA}.
We also find that such an observation is not rare. For instance, according to the statistics in Fig.~\ref{fig:impact} (Adult), roughly $470$ ($1.5\%$ of the training dataset) unique entries at least double their exposure; roughly $150$ ($0.47\%$ of the training dataset) unique entries at least triple their exposure.

Next, we explore the factors that potentially trigger our observations by a set of experiments inspired by unintended memorization~\cite{Carlini2019Secret}. Specifically, unintended memorization identifies the impact of the presence of one training input on the modeled distribution $\mathbf{P}_g$ learned by GAN. Note that this experiment resembles the definition of differential privacy (DP)~\cite{dwork2006calibrating}. DP is more generic and rigorous as it measures that the probability difference varies all possible functions and all data points, which is  computationally infeasible in our measurements. In this case, we narrow down the design by observing the difference between a sample density in two generated distributions $\mathbf{P_{g}}$ trained on neighboring training sets.

\begin{figure}[!t]
	\centering  
	\includegraphics[width=\columnwidth]{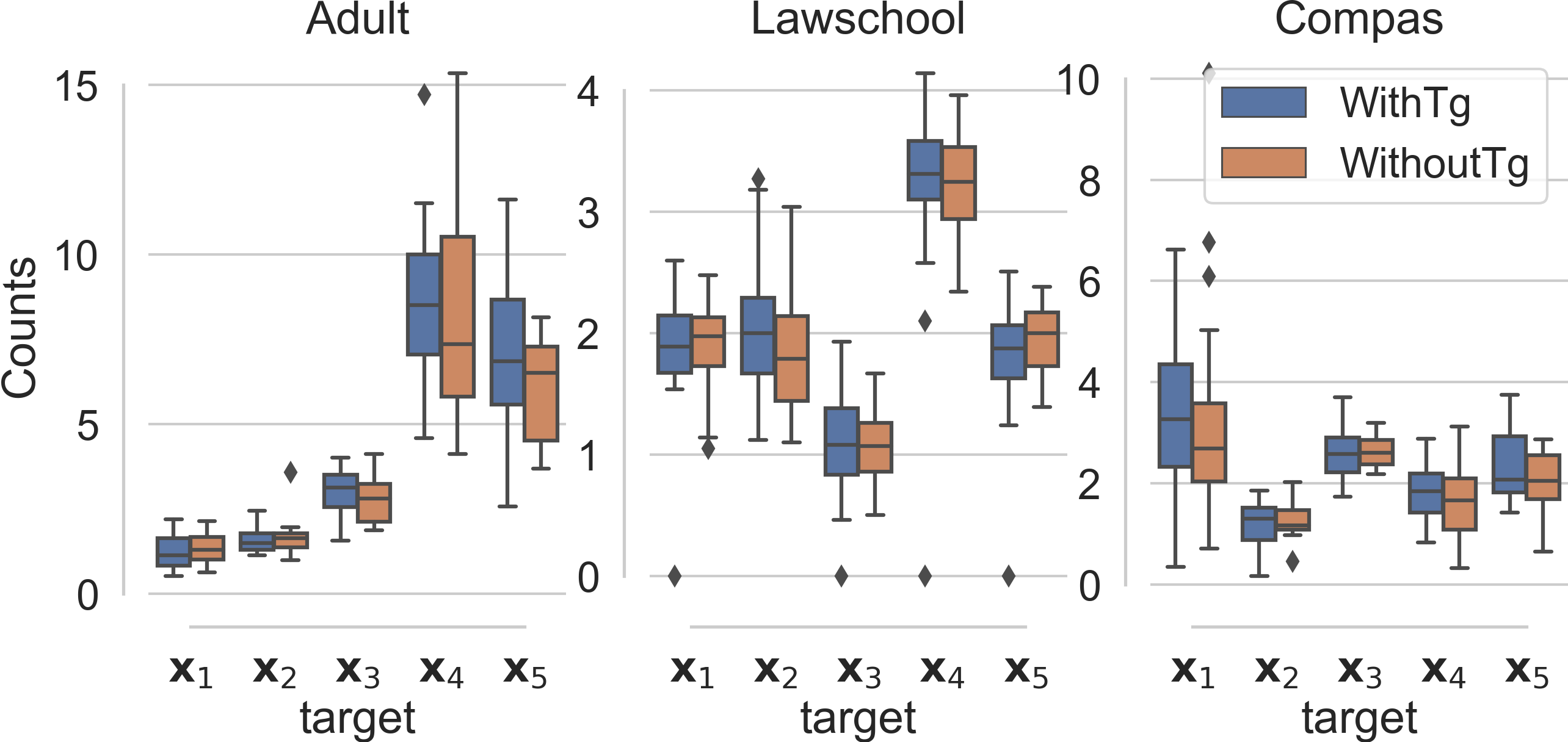}
	\vspace{-3mm}
	\caption{Memorization experiments on three datasets. The blue-color boxplot depicts the frequencies when the target entry is in the training set while the orange one depicts the frequencies when the target entry is deleted from the training set.}
	\label{adult_counts}
\end{figure}

Let $D_t$ be the sensitive training set, $\mathbf{x}_i \in D_t$ be a target data point and $D^{\prime}_{t} = D_t^{\setminus \mathbf{x}_i}$ be the neighboring dataset such that the Hamming distance ${\rm{d_H}}(D_t,D^{\prime}_{t})=1$. Let $G$ be a learned generator trained on $D_t$ and $G^{\prime}$ be a generator trained on $D^{\prime}_{t}$. We measure the difference between the probability of producing a synthetic data point $\mathbf{x}_i$ with (prior) and without (posterior) the input data point~$\mathbf{x}_i$. 
\begin{equation}
\label{pro}
\vspace{-1mm}
	\frac{\Pr(G(z)=\mathbf{x}_i \mid D_t)}{\Pr(G^{\prime}(z)=\mathbf{x}_i \mid D^{\prime}_{t})}
\vspace{-1mm}
\end{equation}
Following a recent work~\cite{Carlini2019Secret}, GAN models do not memorize a data point if it does not exist in the training dataset $D_t$. Thus, if Eq.~\eqref{pro} approaches  $1$, we say that the target data $\mathbf{x}_i$ is unlikely to be memorized by the GAN. 
The pseudo-code of the experiment is presented in Alg.~\ref{alg:memo}. In the experiment, we use $20$ different GANs ($N_k = 20$) and some of target data  to estimate Eq.~\eqref{pro}. We report the experimental results of five target data ($N_c=5$) in Fig.~\ref{adult_counts}. 

\begin{algorithm}[t]
\DontPrintSemicolon
\KwInput {
$\{\mathbf{x}_{1}, \dots, \mathbf{x}_{N_c}\}$: sample data points; $D_{t}$: private training dataset.
} 
\KwOutput{$\{\Pr[\mathbf{x}_k|G_i(z)]\}$: prior frequency; $\{\Pr[\mathbf{x}_k|G^{\prime}_i(z)]\}$: posterior frequency.}
\While {$k: 1 \to N_{c}$}
{
    \While {$i: 1 \to N_{k}$}
    {
	   Generative model $G_{i}$ $\gets$ Train on $D_{t}$;\;
	   $\Pr[\mathbf{x}_k|G_i(z)]$ $\gets$ Estimate by Eq.~\eqref{eq:prob};\;
	   $D^{\prime}_{t}$ $\gets$ $D_{t} \setminus \{\mathbf{x}_{k}$\};\;
	   Generative model $G^{\prime}_{i}$ $\gets$ Train on $D^{\prime}_{t}$;\;
	   $\Pr[\mathbf{x}_k|G^{\prime}_i(z)]$ $\gets$ Estimate by Eq.~\eqref{eq:prob};\;
	}
}
\Return{$\{\Pr[\mathbf{x}_k|G_i(z)]\}$, $\{\Pr[\mathbf{x}_k|G^{\prime}_i(z)]\}$}
\caption{Memorization Experiment.}
\label{alg:memo}
\end{algorithm}

From Fig.~\ref{adult_counts}, we choose the same samples (data points) as in Fig.~\ref{fig:impact} to compare how prior (with a target $\mathbf{x}_i$) and posterior densities (without target $\mathbf{x}_i$) differ in modeled distribution. We find that the presence of the target entry $\mathbf{x}_i$ has limited influence on its frequency in modeled distribution $\mathbf{P}_g$. Even if some data point $\mathbf{x}_i$ is absent in the training set, its probability density in synthetic distribution $\mathbf{P}_g$ is still high, e.g., $\mathbf{x}_4, \mathbf{x}_5$ in the Adult dataset. 
This is perhaps because GAN's generalization smooths the sudden change that happened in the probability space of the training set. For instance, the density of the a target point $\mathbf{x}_i$ in $\mathbf{P}_r$ may be lower than the surrounding points, whereas the GAN smooths such sudden changes in the probability space, and thus it is unintended to increase its probability of exposure. In another aspect, such a rough probability space in real distribution may be attributed to insufficient sampling or unbalanced sampling. As such, cautious data collection may have positive impact in mitigating such influence. Understanding this complicated phenomenon with more explicit proof is our future work. Currently, we summarize that the unique training data recovered by \emph{TableGAN-MCA} is mainly due to the GAN's generalization rather than the unintended memorization. This result implies that mitigating the attack effect of \emph{TableGAN-MCA} may inevitably compromise the availability of released synthetic datasets, since GAN generalization is closely related to its generation ability, which potentially impacts the quality of generated data. 

\section{Mitigation}
In this section, we evaluate the mitigation effects of differential privacy and two customized defense methods against \emph{TableGAN-MCA}.

\subsection{Differentially Private WGAN-WC}
\label{DP_sec}
\begin{figure*}[!t]
\centering  
\subfigure[Adult]{\includegraphics[width=0.33\textwidth]{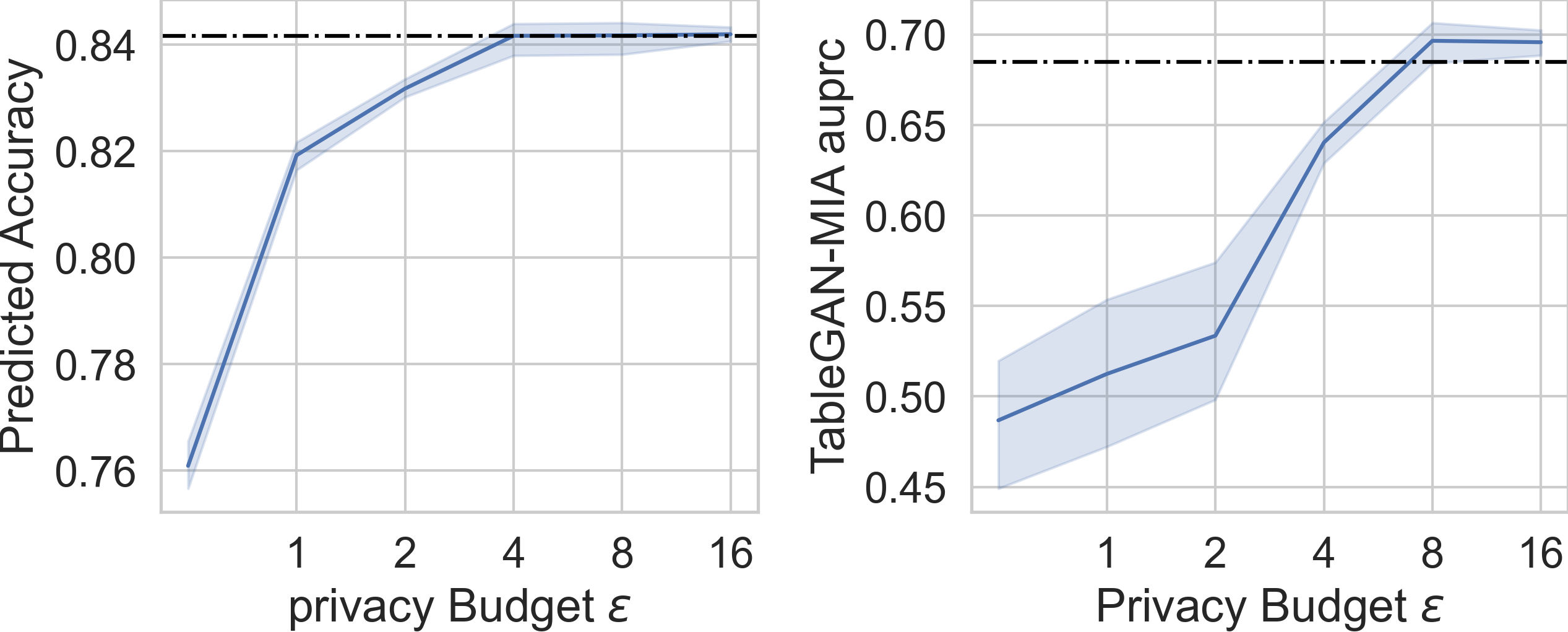}}
\subfigure[Lawschool]{\includegraphics[width=0.33\textwidth]{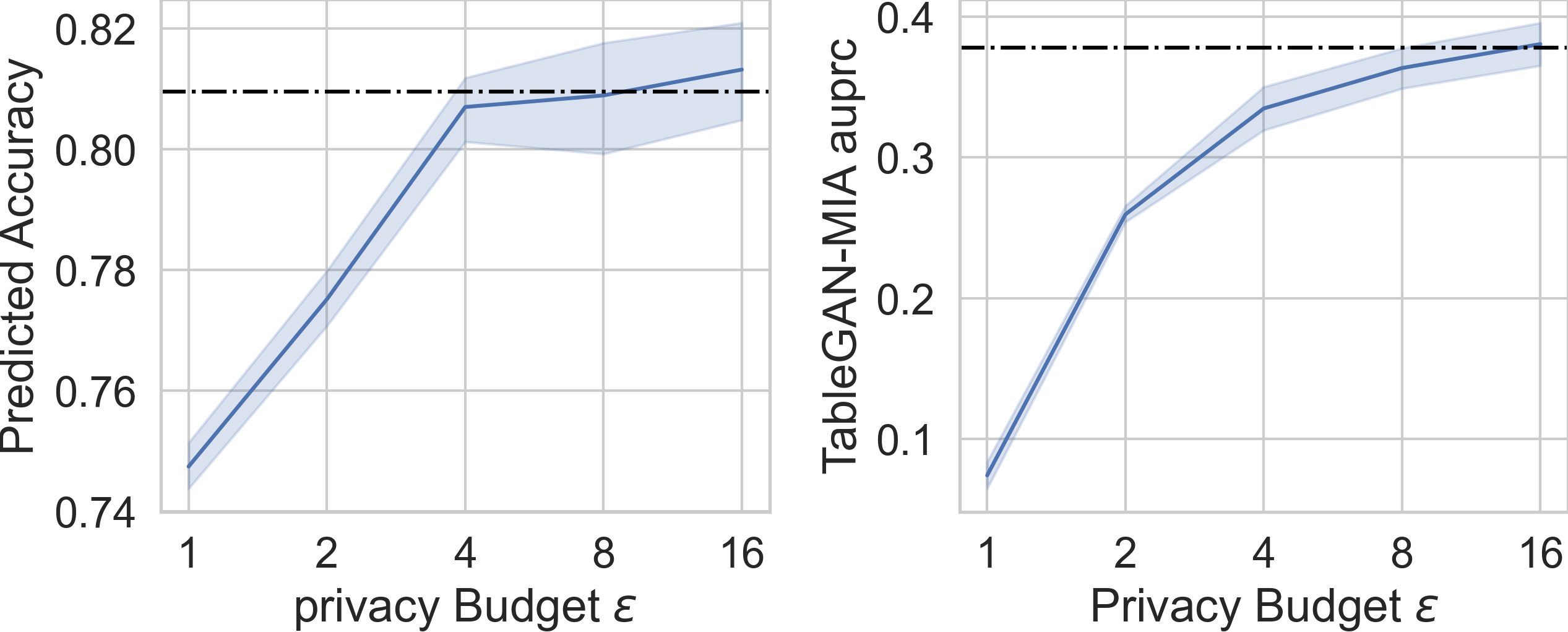}}
\subfigure[Compas]{\includegraphics[width=0.33\textwidth]{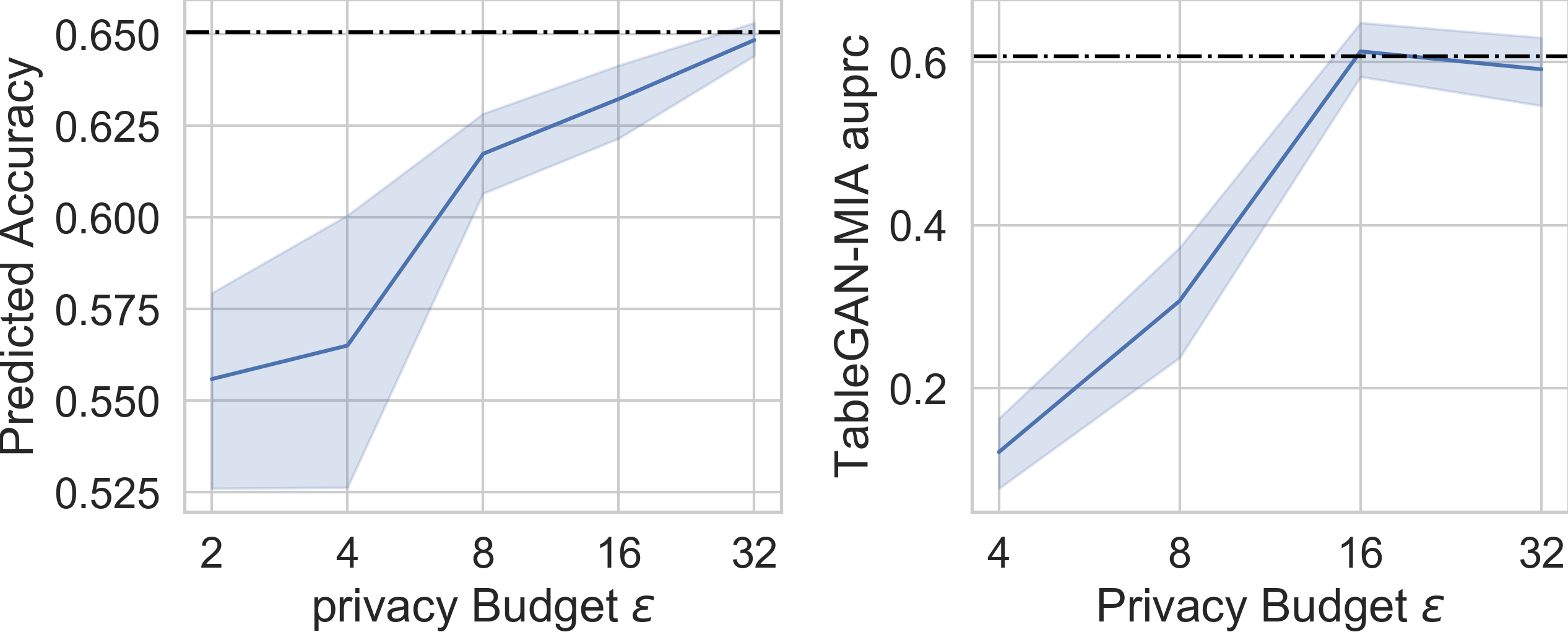}}%
\vspace{-3mm}
\caption{Differential private GAN-synthesized data utility (left) and \emph{TableGAN-MCA} effect (right) for Adult, Lawschool and Compas benchmarks. Dash dot line denotes non-private WGAN with weight clipping baseline.}
\label{fig:dp}
\end{figure*}

\textbf{Differentially Private WGAN (DP-GAN) only has acceptable trade-offs for larger privacy budgets, and may hardly eliminates \emph{TableGAN-MCA} without compromise synthetic data utility.}
Differential privacy~\cite{dwork2006calibrating} provides a quantified solution to output randomized answers. In this work, we apply a standard approach of differentially private iterative training procedure (DP-SGD, short for DP stochastic gradient descent)~\cite{abadi2016deep,mcmahan2018general} to the GAN to train a ($\epsilon, \delta$)-differentially private generator oracle. Otherwise, since DP-SGD perturbs the training process of discriminative models, such mitigation may achieve sub-optimal trade-offs between membership collision privacy and synthetic data utility. 
 In the experiments, we implement the DP framework according to \cite{mcmahan2018general} and account the privacy budget $(\epsilon,\delta$) using RDP accountant released in Tensorflow/Privacy project.  Note that  WGAN-GP, TVAE and CTGAN do not have DP versions, and thus we study the DP version of WGAN-WC. The generation quality and \emph{TableGAN-MCA} effect of non-private baseline are shown in Fig.~\ref{fig:dp} followed by Table~\ref{tab:ML} and Fig.~\ref{fig:auprc_targets}.

To implement DP-WGAN, we train a differentially private discriminator. The generator is differentially private because of the post-processing~\cite{dwork2014algorithmic}. We add calibrated noise into each gradient of the discriminator during training. The accumulation of multiple Gaussian noise addition~\cite{dwork2006our} relies on privacy accountant techniques~\cite{abadi2016deep} and R{\'e}nyi differential privacy~\cite{mironov2017renyi}. We provide DP-related hyper-parameters in Table~\ref{Hyper}, Appendix~\ref{DP_details}.

We provide the experimental results of the machine learning utility and \emph{TableGAN-MCA} effect when sharing differentially private synthetic data in Fig.~\ref{fig:dp}. The shadow GANs in use are non private WGAN-WC. The privacy budget $\epsilon$ measures the amount of privacy leakage and a smaller value means more privacy-preserved. $\delta$ denotes the probability of violating $\epsilon$-DP, which is set to $\frac{1}{O(|D_t|)}$. As can be seen from Fig.~\ref{fig:dp}, the DP method has some positive effect in defending against the \emph{TableGAN-MCA}. For Adult datasets, when privacy budget $\epsilon \approx 2.0$, the attack AUPRC decreases by $16.01\%$ and model's predicted accuracy decreases by $1.18\%$ in comparison to the no-DP baseline (see dash dots in Fig.~\ref{fig:dp}). For the Compas dataset, when privacy budget $\epsilon \approx 8.0$, the attack AUPRC decreases by $48.33\%$ and model's predicted accuracy decreases by $5.13\%$ in comparison to the no-DP baseline. We also depict the ECDF comparison between the original training data and differentially private synthetic data for each marginal to show marginal fitness compromise in Fig.~\ref{fig:dp_ecdf} (Appendix~\ref{DP_details}).  It is not surprising that DP-WGAN achieves sub-optimal trade-offs when protecting against \emph{TableGAN-MCA}, since the memorization experiment shows that the presence of individuals does not significantly affect the generated distribution. The membership collisions information that we intend to infer is perhaps highly correlated to population statistics (attributes correlation), which will be preserved even under DP training.

 \begin{figure}
 \centering  
 \subfigure[No Defense]{
 \includegraphics[width=0.36\columnwidth]{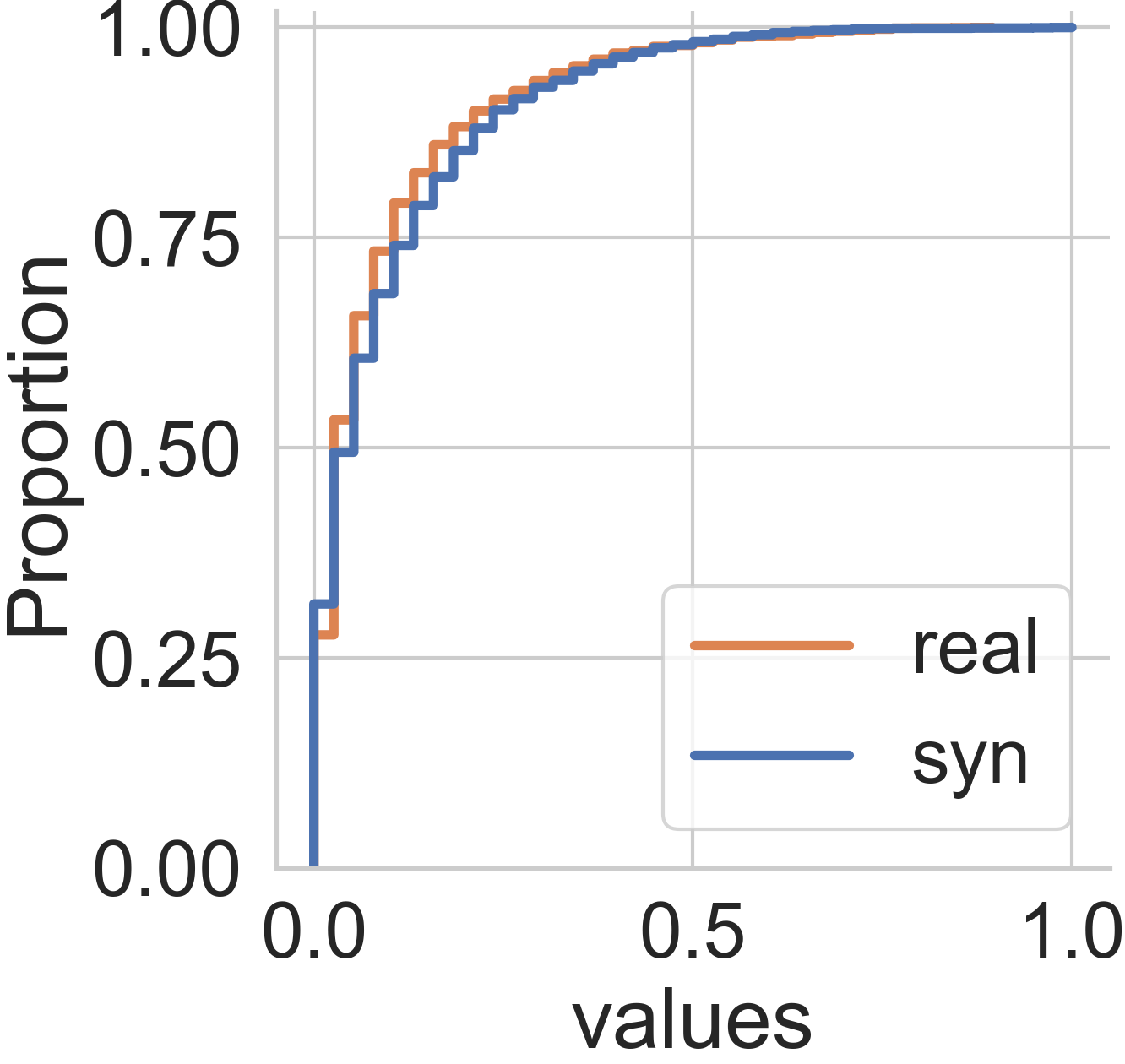}
 \label{no_defense}
}%
 \subfigure[Naive Defense]{
 \includegraphics[width=0.28\columnwidth]{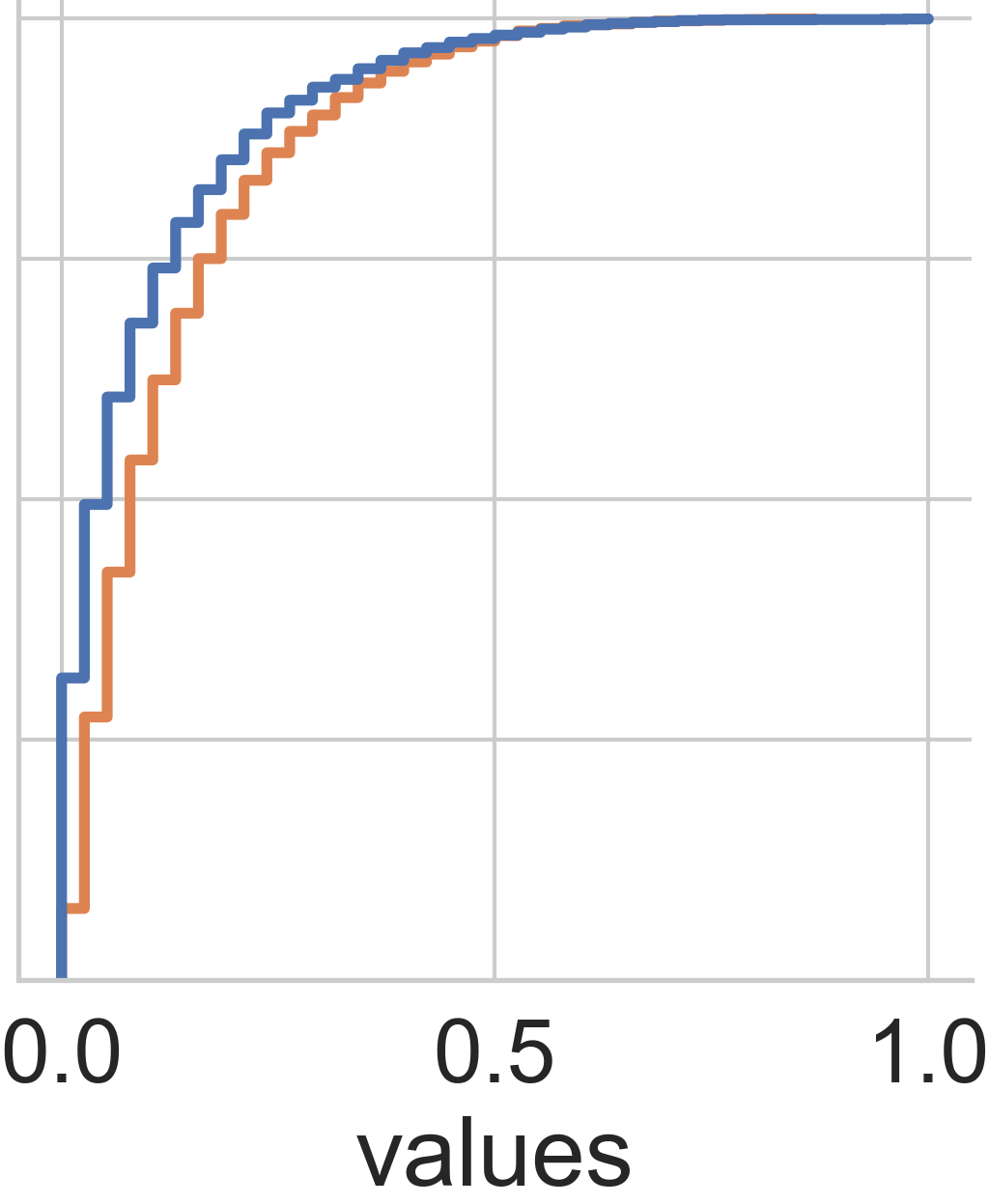}
 \label{naive_defense}
}
 \subfigure[Improved Defense]{
 \includegraphics[width=0.28\columnwidth]{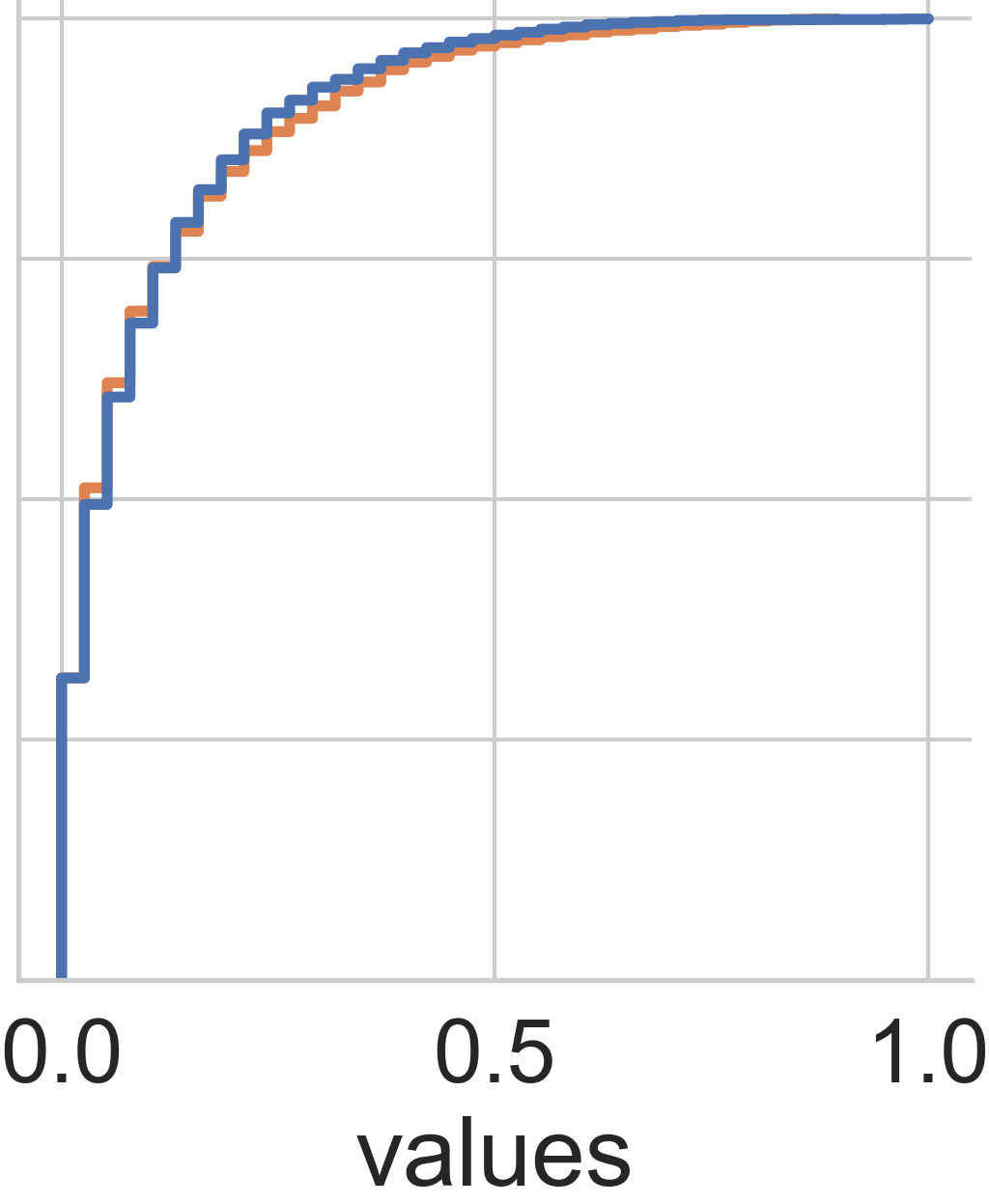}
 \label{improved_marginal}
}
 \vspace{-3mm}
 \caption{ECDF comparisons for synthetic datasets generated by three methods. We choose ``priors count'' attribute in the Compas dataset.}
 \label{ECDF_compare}
 \end{figure}

\subsection{Customized Defense}
\subsubsection{Remove Colliding Members}
\textbf{Removing colliding members protects against \emph{TableGAN-MCA} but it reduces the distribution fitness.} The straightforward solution against  \emph{TableGAN-MCA} is to manually remove colliding members from the sampled synthetic dataset and share a cleaned version to the analysts (customers). The whole process is denoted as the ``naive defense'' (last steps in Alg.~\ref{alg:improved}). We acknowledge the cleaned version can decrease the utility of original synthetic data, especially for distribution fitness. For example, we present the ECDF comparison of synthetic datasets generated by the naive defense (Fig.~\ref{naive_defense}) and no-defense  (Fig.~\ref{no_defense}). We show that the naive defense exhibits  decreased marginal fitness compared with no-defense baseline. More ECDFs can be found in Figs.~\ref{fig:naive_adult}, ~\ref{fig:lawsch_naive}, and~\ref{fig:compas_naive} (Appendix~\ref{NaiveandImprovedDefenses}).

\subsubsection{GAN-constrained Training}
We propose a \textbf{GAN-constrained training technique}, to further improve synthetic data utility while protecting against \emph{TableGAN-MCA}. This strategy is denoted as an ``improved defense''. Simply put, we motivate GANs to generate a synthetic dataset $S \sim \mathbf{P}_g$ that is disjoint with the training set $D_t$ while minimizing the distance between training data and generated data $\mathcal{L}(D_t,S)$, which is 

 \begin{equation}
 S=\operatorname{arg}\underset{S_i}{\min}~\mathcal{L}(S_i,D_t)|_{S_i\cap D_t = \varnothing}, 
 \end{equation}
 
 where $\mathcal{L}$ denotes a distance metric. Since the discriminator of the WGAN minimizes the Wasserstein distance, we additionally add a constraint during training to force each sampled batch of the generator to be disjoint with $D_t$. To do so, we remove the intersection between the sampled batch and the training set every iteration before computing the loss function (see Alg.~\ref{alg:improved}). Thus, WGAN  automatically searches for the best substitution for such  colliding samples at training.

\begin{algorithm}[t]
\DontPrintSemicolon
\KwInput {
$D_t$: private training data; $N_g$: number of discriminator iterations per generator iteration; $m$: batch size
} 
\KwOutput{A Synthetic dataset $S$}
\For {each iteration}
{
     \While {$i: 1 \to N_{g}$}
     {
	   Sample $\left\{\mathbf{x}^{(i)}\right\}_{i=1}^{m} \sim \mathbf{P}_r$;\;
	   Sample $\left\{\mathbf{z}^{(i)}\right\}_{i=1}^{n} \sim \mathbf{P}_{\mathbf{z}}$, $n>m$; Choose $m$ of $n$ priors $\left\{\mathbf{z}^{(i)}\right\}_{i=1}^{m}$ s.t., $G(\mathbf{z}) \not\in D_t$\;
	   Compute loss, backward, update gradients;\;
	   }
	   Sample $\left\{\mathbf{z}^{(i)}\right\}_{i=1}^{n} \sim \mathbf{P}_{\mathbf{z}}$; Choose $m$ of $n$ priors $\left\{\mathbf{z}^{(i)}\right\}_{i=1}^{m}$ s.t., $G(\mathbf{z}) \not\in D_t$;\;
	   Compute loss, backward, update gradients;\;
}
$S \gets G(\mathbf{z}), s.t., G(\mathbf{z}) \not\in D_t$; \Comment{Naive defense} \;
\Return{$S$}
\caption{GAN-constrained Training (Improved defense)}
\label{alg:improved}
\end{algorithm}

\subsubsection{Naive and Improved Defenses Evaluation}  
\begin{figure}[!t]
 \centering 
 \subfigure[ML Prediction]{
 \includegraphics[width=0.48\columnwidth]{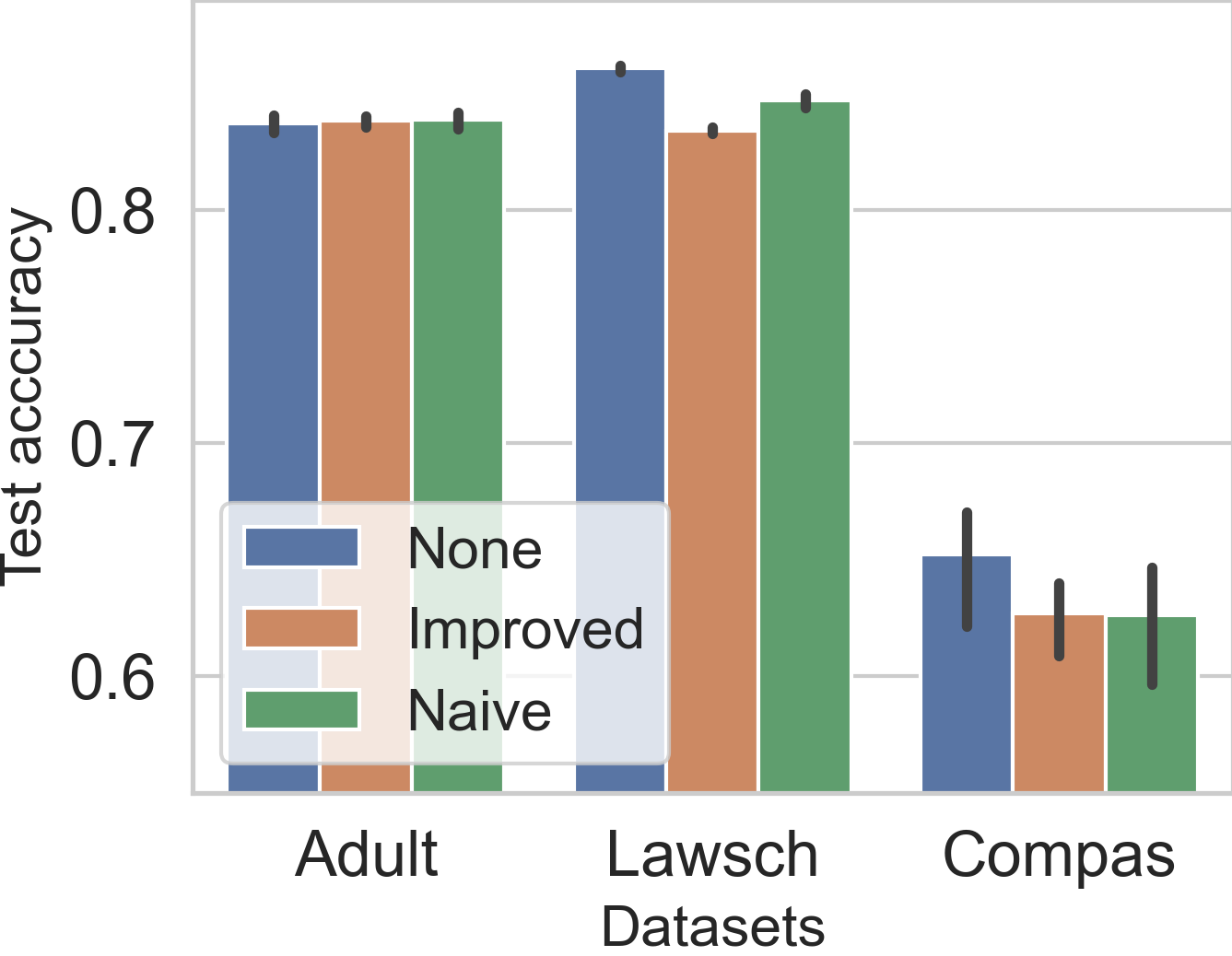}
  \label{defense_a}}%
 \subfigure[Marginal Likelihood]{
 \includegraphics[width=0.48\columnwidth]{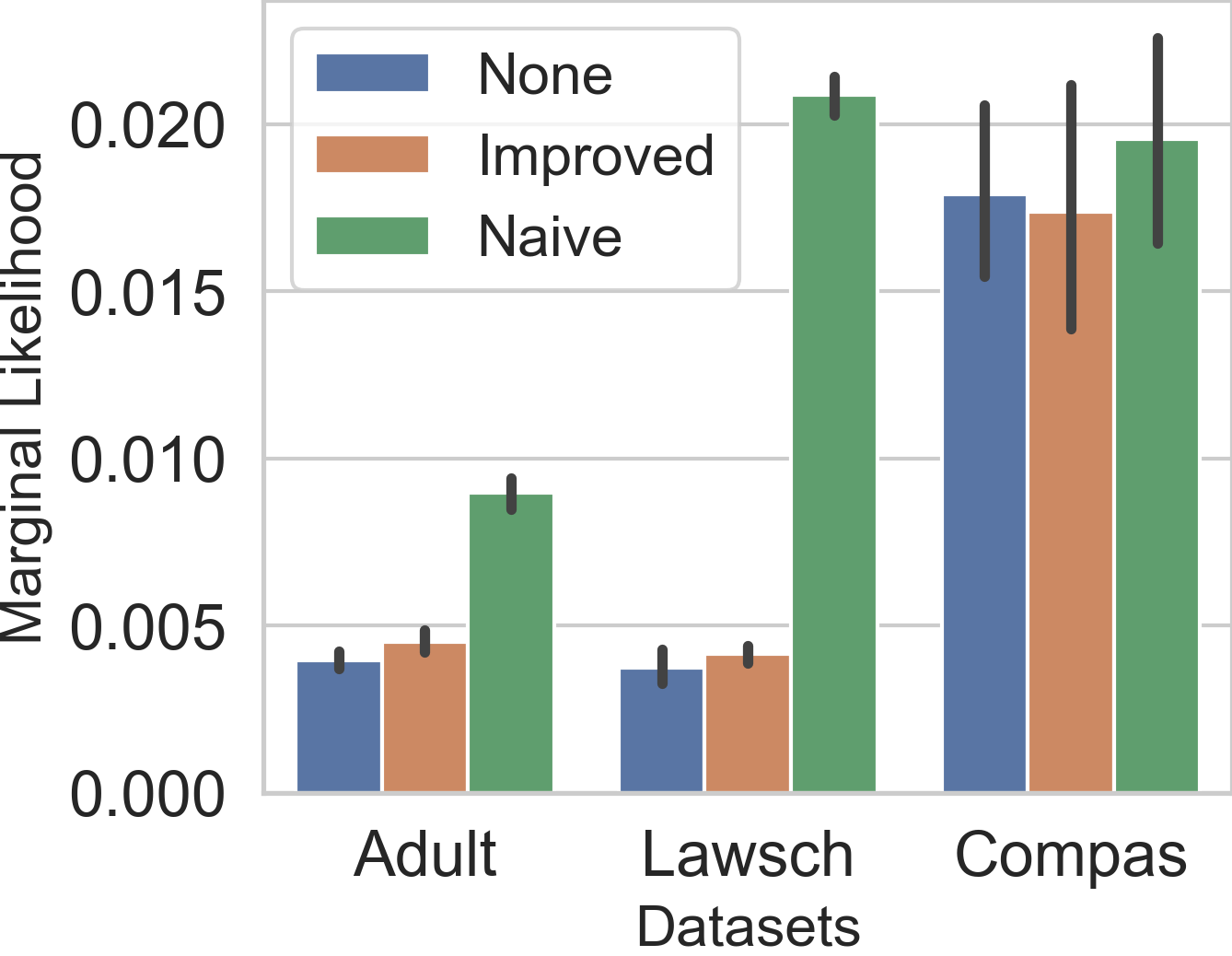}
  \label{defense_b}}
 \vspace{-3mm}
 \caption{Defenses comparisons on three datasets. Marginal Likelihood: compute $\mathbb{E}_i (l_{1})$.}
 \label{Defense_compare}
 \end{figure}

\textbf{The improved defense in large part achieves superior trade-offs than the naive defense, and is almost comparable to the no-defense baseline.} We evaluate synthetic data utility of the naive defense, the improved defense and the no-defense (baseline) on WGAN-GP. Note that the baseline is vulnerable to \emph{TableGAN-MCA} while naive and improved defenses protect against it. We evaluate machine learning efficacy in Fig.~\ref{defense_a} and marginal fitness in Fig.~\ref{defense_b}.
 
In Fig.~\ref{defense_a}, we train machine learning models (Logistic Regression Classifier) on synthetic data sampled from the naive defense, the improved defense and the  no-defense generator and predict on the real test data. Fig.~\ref{defense_a} shows that synthetic data generated by naive and improved defenses achieve satisfying prediction accuracy on the Adult and Lawschool datasets. In the Compas dataset, mitigation methods decrease the prediction accuracy compared to the no-defense baseline.

In Fig.~\ref{defense_b}, we compare ECDFs using $\mathbb{E}_i (l_{1})$ (recall Section~\ref{utility_metrics}). The lower score implies better marginal fitness. The experimental result shows that the improved defense outperforms  the naive defense, and is on par with the no-defense baseline. The improved defense succeeds in compensating the statistical deviation caused by the naive defense (see Fig.~\ref{improved_marginal}). More ECDFs of the naive defense and the improved defense are shown in Figs.~\ref{fig:naive_improved_ecdf_adult},~\ref{fig:naive_improved_ecdf_lawsch}, and~\ref{fig:naive_improved_ecdf_compas}. 

In summary, both naive and improved defenses protect against \emph{TableGAN-MCA} and in part preserve learning ability of released synthetic data. Moreover, the improved defense achieves better marginal fitness than the naive defense. Despite the potentially effective mitigation, \emph{TableGAN-MCA} still remains a threat since the proposed defenses achieve sub-optimal privacy-utility trade-offs, eg, reduced synthetic data diversity, under-performance for tiny-domain datasets (see Compas datasets for details).

\section{Related Work}
\label{sec:related_work}

Membership privacy is the existence of individuals~\cite{li_membership_2013, rahman_membership_2018}. Existing studies show membership disclosure on discriminative machine learning models, e.g., classifiers~\cite{shokri2017membership,yeom2018privacy,long2018understanding,salem_ml-leaks_2018,nasr2019comprehensive,pmlr-v97-sablayrolles19a} and generative machine learning models, e.g., Generative Adversarial Networks~\cite{park2018data,hayes_logan_2019,hilprecht_monte_2019,chen_gan-leaks_2020}.
In the discriminative settings, an adversary infers whether a specific data point is used to train a target model by querying classifier APIs and using predicted probability vectors, labels, logits, etc., to train attack models. For instance, Shokri's shadow model~\cite{shokri2017membership} infers membership against overfitted multi-class classifiers by training an attack model with labeled synthetic data, which mimic the private training data. Subsequent works further relax the adversary's background knowledge~\cite{salem_ml-leaks_2018} by extending attacks to the white-box~\cite{nasr2019comprehensive} and the label-only settings~\cite{li_label-leaks_2020,choo_label-only_2020}. 

In the track of inferring membership against the generative models, there are several successful approaches, such as, 
table-GAN~\cite{park2018data}, LOGAN~\cite{hayes_logan_2019}, MC~\cite{hilprecht_monte_2019} and  GAN-leaks~\cite{chen_gan-leaks_2020}. Note that some of these approaches is originally proposed against image data; however, they are possibly extendable to attack tabular data. That is, they are all related to this study. Hence, we briefly summarize these methods in this section. The conceptual comparisons are shown in Table~\ref{tab:compare}. 
LOGAN~\cite{hayes_logan_2019} and table-GAN~\cite{park2018data} leverage the output of the overfitting discriminator to train an attack model, which is a variant of Shokri et al.~\cite{shokri2017membership} in the context of GAN synthesis. However, their attacks require the predicted probability vector of the target discriminator at the inference phase (see column $3$ in Table~\ref{tab:compare}). In our experiment, we have already shown that a GAN resilient to their attacks may still expose training data to \emph{TableGAN-MCA}. 
MC~\cite{hilprecht_monte_2019} and GAN-leaks~\cite{chen_gan-leaks_2020} extract a customized membership indicator of an overfitting generator to train an attack model. We share a similar theoretical bases with theirs, that is, the modeled distribution of the generator behaves differently on training input versus the non-training one. However, our attack further recovers partial training data by inferring membership of published synthetic data, which is out of their scope (see Columns $4$ and $5$ in Table~\ref{AP_compare}). In this work, we empirically show that the membership inference classifier cannot be directly used to identify membership collisions in our attack model (see Table~\ref{AP_compare}). Compared to those works, we propose a novel attack model, \emph{TableGAN-MCA}, that exposes partial training data by exploiting the weakness of tabular data synthesis. Even though we share similar ideas with MIAs in generative setting, the attack model of \emph{TableGAN-MCA} learns different decision boundaries. According to the experimental results, the success of the proposed attack relies more on population knowledge than individual presence, which is different from MIAs.

\section{Conclusion}

GAN-synthesized table releasing provides unprecedented opportunities for private data sharing that aims to study the regular pattern of population. 
In this work, we propose a novel membership collision attack, \emph{TableGAN-MCA}, against the GAN-synthesized table. Our comprehensive experiments over the real-world datasets conclude some important findings. \emph{TableGAN-MCA} achieves high recovering rate against the private training data from the published GAN-synthesized tables. Our in-depth studies suggest that the target model, training data size, training epochs and training data frequencies impact the attack performance of \emph{TableGAN-MCA}. We further conclude that the training data leakage is mainly related
to the published population statistics (attributes correlations), rather than the model memorization. To mitigate the effect of \emph{TableGAN-MCA}, we find that differential privacy (applying DP-WGAN) does not show a satisfying result mainly due to the correlations between training data features. Based on our understanding on \emph{TableGAN-MCA}, we propose two mitigation approaches, which substitute the published colliding members with similar non-private data entries.
We hope that the concept of membership collisions defined and the attack methodology developed in this paper could inform the privacy community of such new potential leakage of data synthesis.

\begin{acks}
The authors, affiliated with Southeast University, were partially supported by Jiangsu Provincial Key Laboratory of Network and Information Security (No. BM2003201). Minhui Xue was, in part, supported by the Australian Research Council (ARC) Discovery Project (DP210102670). Aiqun Hu and Minhui Xue are the corresponding authors of this paper. 
\end{acks}

\bibliographystyle{ACM-Reference-Format}
\bibliography{main}

\appendix
\section*{Appendix}

\section{Network Structure and Parameters}

\label{app_a}

WGAN-GP, WGAN-WC shares the same network architecture. We set the Generator as Recurrent Neural Networks (RNNs). According to our experiments, the RNN has a positive effect on stabilizing the generator's outputs. Eq.~\eqref{equa4} represents the Generator networks and Eq.~\eqref{equa5} represents the Discriminator networks. 

\begin{equation}
\label{equa4}
\left\{
\begin{array}{lr}
h_1 = \textsf{ReLU}(\textsf{BN}(\textsf{FC}_{|z| \rightarrow 256}(z))) \\
h_2 = \textsf{ReLU}(\textsf{BN}(\textsf{FC}_{|z|+256 \rightarrow 256}(z\oplus h_1)))\\
G(\cdot)_{con}= \textsf{gumbel}_{0.2}(\textsf{FC}_{|z|+512 \rightarrow |r|}(h_2))   \\
G(\cdot)_{cat}= \textsf{tanh}(\textsf{FC}_{|z|+512 \rightarrow 1}(h_2))
\end{array}
\right.
\end{equation}

\begin{equation}
\label{equa5}
\left\{
\begin{array}{lr}
h_1 = \textsf{dropout}_{0.5}(\textsf{leakyReLU}_{0.2}(\textsf{FC}_{|r| \rightarrow 256}(r))) \\
h_2 = \textsf{dropout}_{0.5}(\textsf{leakyReLU}_{0.2}(\textsf{FC}_{256 \rightarrow 256}(h_1)))\\
D(\cdot)= \textsf{FC}_{256 \rightarrow 1}(h_2)) &  
\end{array}
\right.
\end{equation}

For TVAE and CTGAN, we applies the module \verb|CTGANSynthesizer| and \verb|TVAESynthesizer| of the \verb|SDGym|~\cite{SDGym}. Thus, the structures and hyper-parameters are exactly same as the originals'~\cite{xu2019modeling}.

\noindent \textbf{Hyper-parameters.} For Adult and Lawschool datasets, we train $300$ epochs and set batch size to $500$. For Compas dataset, we train $600$ epochs and set batch size to $100$. Since the Compas dataset is much smaller than others, we find that less iterations could incur under-fitting. Additionally, balancing the number of D and G training sessions also helps to converge faster.

\section{Number of Synthetic Queries} \label{LSQ}
We thoroughly discuss how the number of synthetic queries influences the attack performance and corresponding attack tricks.

\subsection{Limited Synthetic Queries} 

Many target model prediction APIs (MLaaS) implement a pay-per-query business model. Hence, reducing the number of synthetic queries saves the cost of performing \emph{TableGAN-MCA}. However, a smaller synthetic dataset, having less membership collisions with the training dataset, decays the attack performance. To tackle this problem, we propose an approach that uses shadow data to fill up the synthetic data to match the size of the training set. That is, the adversaries obtain a synthetic dataset $S$ of size $0.25*N$ by querying the target Generator. The adversaries then generate the shadow dataset of size $|\widetilde{S}|=0.75*N$. After that, the \emph{TableGAN-MCA} adversaries attack $S \Vert \widetilde{S}$ instead of the original $S$. 

\begin{figure}
 \centering  
 \includegraphics[width=0.8\columnwidth]{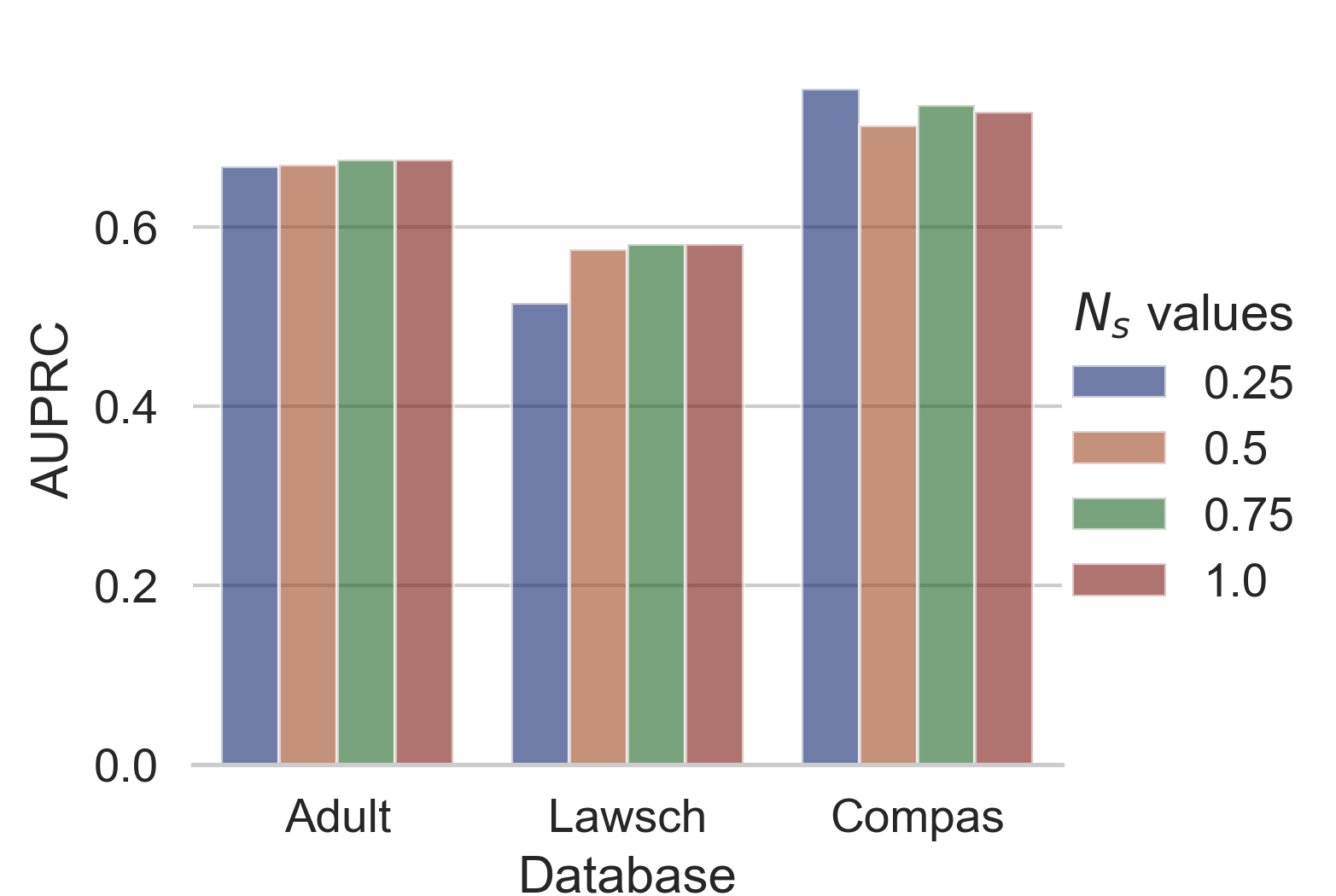}
 \caption{TableGAN-MCA performance when $N_s \leq 1$.}
 \label{few_samplesize}
  \vspace{3mm}
 \end{figure}

We show the impact of a small $N_s$ on \emph{TableGAN-MCA} in Fig.~\ref{few_samplesize}. We find that few synthetic queries also yield decent attack performance. This resonates with the memorization experiment that the success of \emph{TableGAN-MCA} is contingent more on basic data patterns. 

\subsection{Unlimited Synthetic Queries}
The \emph{TableGAN-MCA} adversary continues to expose more training data when increasing the number of synthetic queries. In Fig.~\ref{fig:boost_attack}, we evaluate the \emph{TableGAN-MCA} up to $N_s=10$, which is not the ceiling of \emph{TableGAN-MCA} capabilities. Due to computational constraints, we are limited to performing the attack up to  $N_s=20$ and observe that the number of exposed training data of  \emph{TableGAN-MCA} is still increasing. This leaves open an interesting problem of whether the adversary could reconstruct the whole training dataset with unlimited queries.

\section{Mitigation}
\subsection{DP-WGAN} 
\label{DP_details}

\begin{table}[!t]\small
  \centering
  \caption{Hyper-parameters in DP-WGAN. ($\epsilon$,$\delta$) : privacy budget; $S$: clip threshold; $\sigma$: standard deviation of the noise added in each step. }
  \label{Hyper}%
    \begin{tabular}{cccc}
    \toprule
   Datasets & $(\epsilon,\delta)$ & $(S,\sigma)$ & Sampling rate\\
    \midrule
   \multirow{6}{*}{Adult} 
   & $(0.5,10^{-5})$ &	$(0.1,0.5)$  & 500/31655\\
   & $(1.0,10^{-5})$ &	$(0.1,0.45)$	 & 500/31655\\ 
   & $(2.0,10^{-5})$ &  $(0.1,0.4)$  & 500/31655\\
   & $(4.0,10^{-5})$ &  $(0.1,0.3)$  & 500/31655\\
   & $(8.0,10^{-5})$ &  $(0.1,0.17)$ & 500/31655\\
   & $(16.0,10^{-5})$&  $(0.1,0.11)$ & 500/31655\\
   \midrule
   \multirow{6}{*}{Lawschool}
   & $(0.5,10^{-5})$  &	$(0.1,0.4)$  & 500/43011\\
   & $(1.0,10^{-5})$  &	$(0.1,0.45)$ & 500/43011\\ 
   & $(2.0,10^{-5})$  & $(0.1,0.48)$ & 500/43011\\
   & $(4.0,10^{-5})$  & $(0.1,0.25)$ & 500/43011\\
   & $(8.0,10^{-5})$  & $(0.1,0.15)$ & 500/43011\\
   & $(16.0,10^{-5})$ & $(0.1,0.11)$ & 500/43011\\
   \midrule
   \multirow{6}{*}{Compas}
   & $(2.0,10^{-4})$  & $(0.1, 0.9)$  & 100/3694\\
   & $(4.0,10^{-4})$  & $(0.1, 0.48)$ & 100/3694\\
   & $(8.0,10^{-4})$  & $(0.1, 0.27)$ & 100/3694\\
   & $(16.0,10^{-4})$ & $(0.1, 0.16)$ & 100/3694\\
   & $(32.0,10^{-4})$ & $(0.1, 0.11)$ & 100/3694\\
    \bottomrule
    \end{tabular}%
\end{table}%

\begin{figure*}[t]
\centering  
\subfigure[Adult, $\epsilon = 0.5, \delta = 10^{-5}$]{\includegraphics[width=0.95\textwidth]{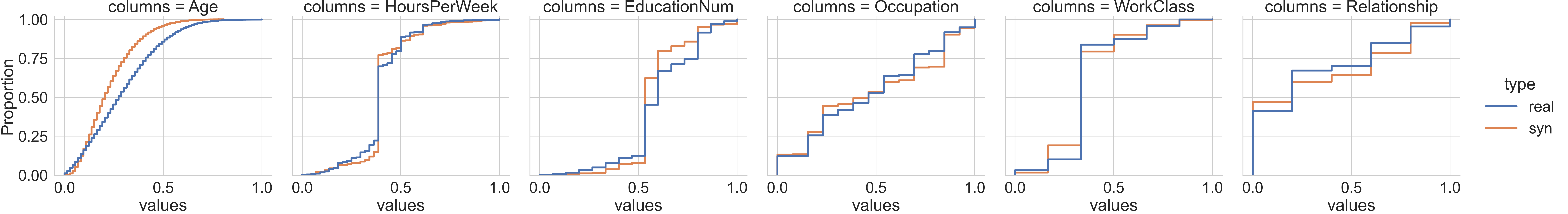}}
\subfigure[Lawschool, $\epsilon = 0.5, \delta = 10^{-5}$]{\includegraphics[width=0.95\textwidth]{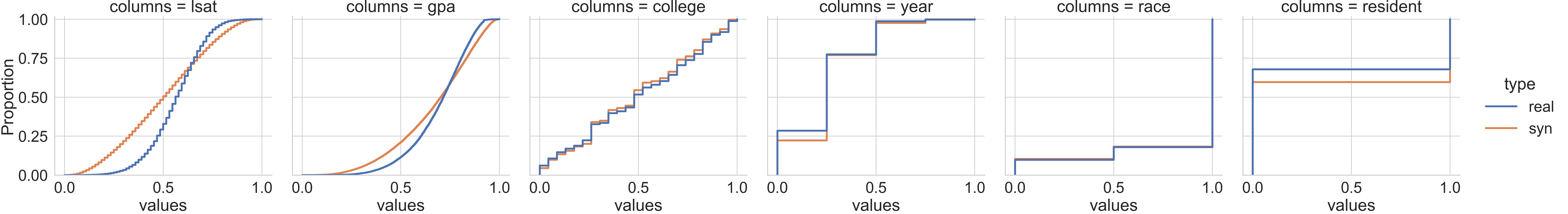}}
\subfigure[Compas, $\epsilon=2.0, \delta=10^{-4}$]{\includegraphics[width=0.95\textwidth]{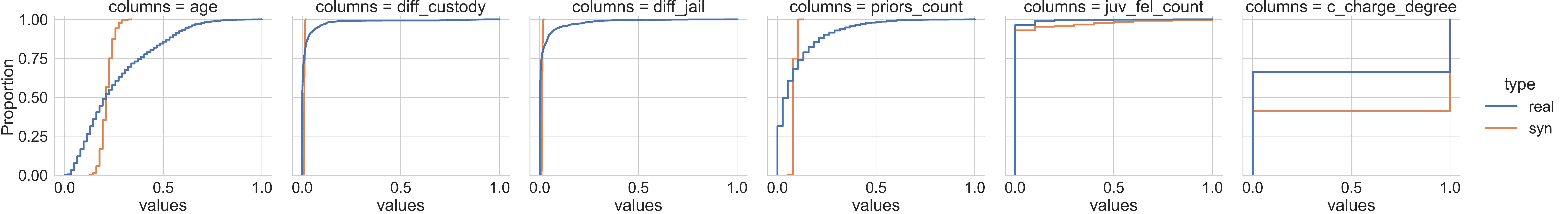}}%
\vspace{-3mm}
\caption{ECDF comparison between training data and differentially private GAN-synthesized data.}
\label{fig:dp_ecdf}
\end{figure*}

We show the ECDFs of marginals for $(\epsilon,\delta)$-DP synthesized data in Fig.~\ref{fig:dp_ecdf}. Smaller training data usually gains less satisfactory generation quality under DP training with a similar privacy budget.

\subsection{Naive and Improved Defenses}
\label{NaiveandImprovedDefenses}

\begin{figure*}
\centering  
\subfigure[Adult ECDF, Remove Overlapping]{\includegraphics[width=0.95\textwidth]{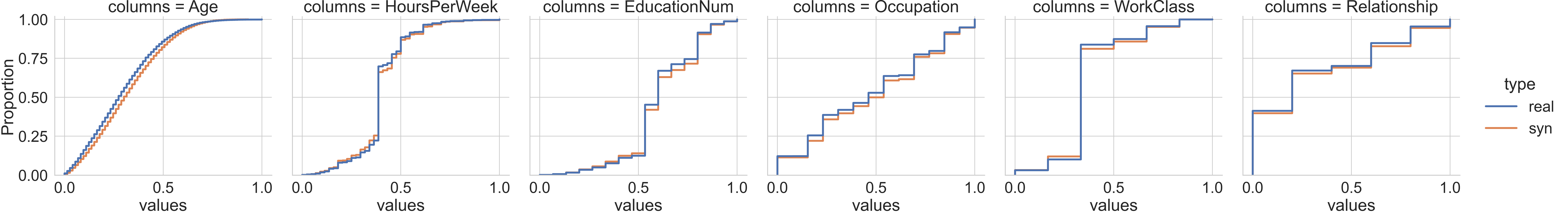}
\label{fig:naive_adult}}
\subfigure[Adult ECDF, GAN-constrained Training]{\includegraphics[width=0.95\textwidth]{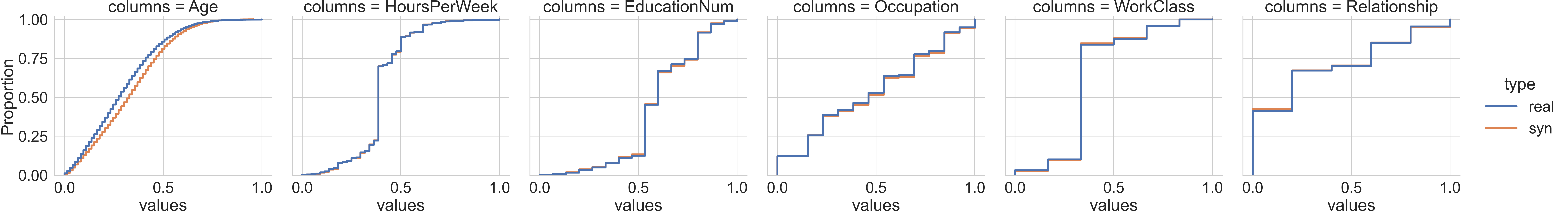}
\label{fig:improved_adult}}
\vspace{-3mm}
\caption{ECDF comparisons for ``Remove Overlapping" mitigation and ``GAN-constrained Training'' mitigation.}
\label{fig:naive_improved_ecdf_adult}
\end{figure*}

\begin{figure*}
\centering  
\subfigure[Lawschool ECDF, Remove Overlapping]{\includegraphics[width=0.95\textwidth]{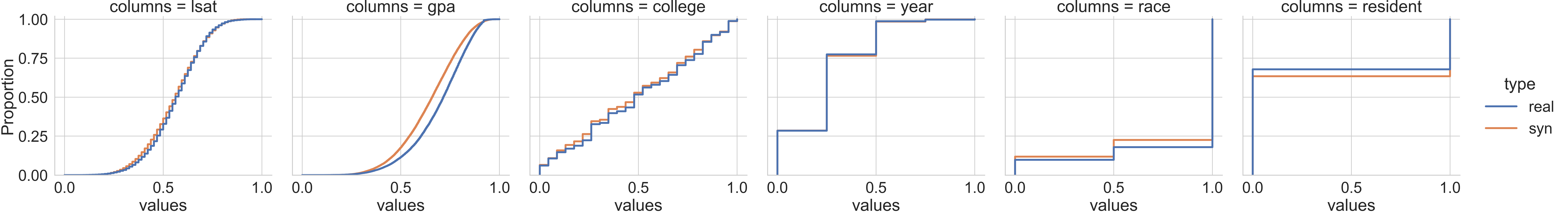}
\label{fig:lawsch_naive}}
\subfigure[Lawschool ECDF, GAN-constrained Training]{\includegraphics[width=0.95\textwidth]{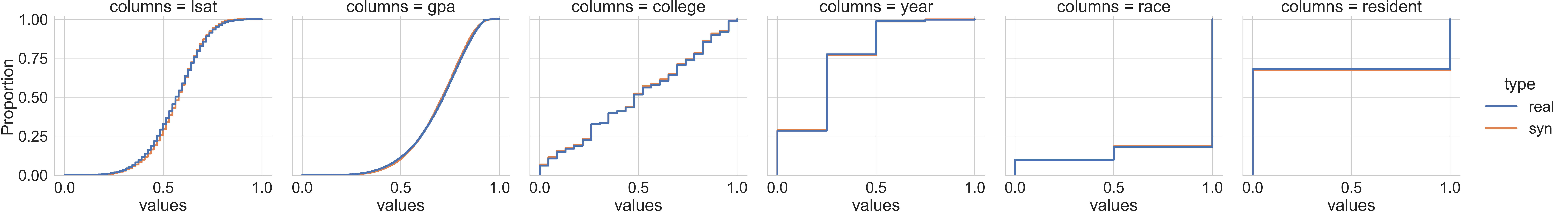}
\label{fig:lawsch_improved}}
\vspace{-3mm}
\caption{ECDF comparisons for ``Remove Overlapping'' mitigation and ``GAN-constrained Training'' mitigation.}
\label{fig:naive_improved_ecdf_lawsch}
\end{figure*}

\begin{figure*}
\centering  
\subfigure[Compas ECDF, Remove Overlapping]{\includegraphics[width=0.95\textwidth]{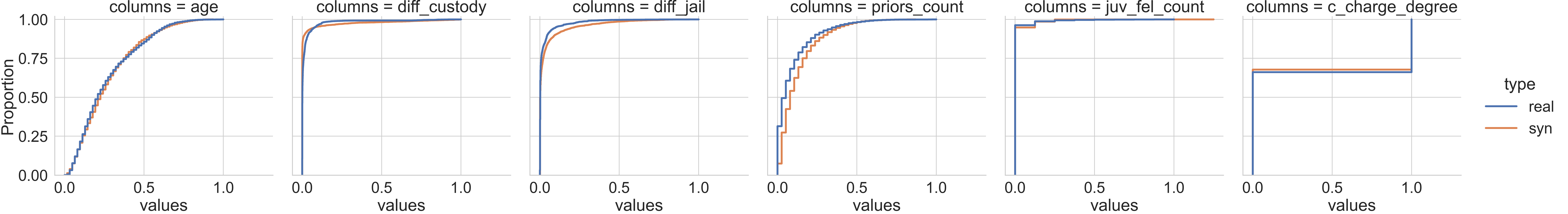}
\label{fig:compas_naive}}
\subfigure[Compas ECDF, GAN-constrained Training]{\includegraphics[width=0.95\textwidth]{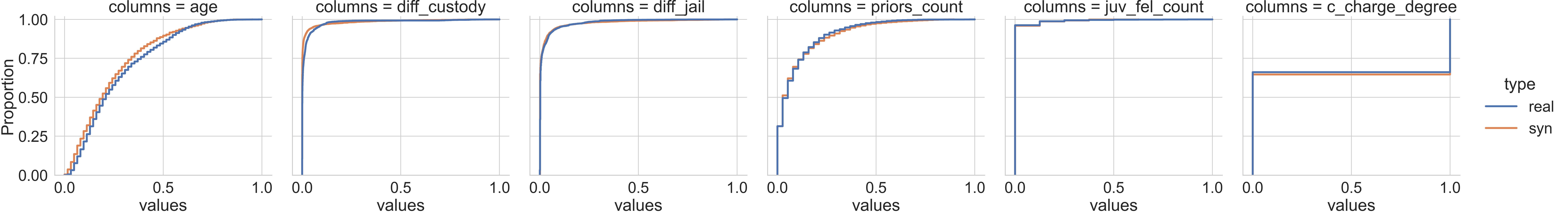}
\label{fig:compas_improved}}
\vspace{-3mm}
\caption{ECDF comparisons for ``Remove Overlapping'' mitigation and ``GAN-constrained Training'' mitigation.}
\label{fig:naive_improved_ecdf_compas}
\end{figure*}

We show additional ECDFs of marginals for ``Remove Colliding Members'' mitigation and ``GAN-constrained Training'' mitigation in Figs.~\ref{fig:naive_improved_ecdf_adult},~\ref{fig:naive_improved_ecdf_lawsch}, and~\ref{fig:naive_improved_ecdf_compas}.

\end{document}